\documentclass[useAMS,usegraphicx,usenatbib]{mn2e}
\usepackage{amssymb}

\title[Disentangling HH~900 in Carina]{Disentangling the outflow and protostars in HH~900 in the Carina Nebula}

\author[Reiter et al.]{Megan Reiter$^{1}$\thanks{E-mail: mreiter@as.arizona.edu}, Nathan Smith$^{1}$, Megan M. Kiminki$^{1}$, John Bally$^{2}$, and Jay Anderson$^{3}$ \\
$^{1}$ Steward Observatory, University of Arizona, Tucson, AZ 85721, USA \\ 
$^{2}$ Center for Astrophysics and Space Astronomy, University of Colorado, 
  Boulder, CO 80309, USA \\
$^{3}$ Space Telescope Science Institute, Baltimore, MD 21218, USA} 

\begin{document}

\date{Accepted 2014 Oct ??. Received 2014 Oct ??; in original form 2014 October ??}

\pagerange{\pageref{firstpage}--\pageref{lastpage}} \pubyear{2014}

\maketitle

\label{firstpage}


\begin{abstract}
HH~900 is a peculiar protostellar outflow emerging from a small, tadpole-shaped globule in the Carina nebula. 
Previous H$\alpha$ imaging with \textit{HST}/ACS showed an ionized outflow with a wide opening angle that is distinct from the highly collimated structures typically seen in protostellar jets. 
We present new narrowband near-IR [Fe~{\sc ii}] images taken with the Wide Field Camera 3 on the \textit{Hubble Space Telescope} that reveal a remarkably different structure than H$\alpha$. 
In contrast to the unusual broad H$\alpha$ outflow, the [Fe~{\sc ii}] emission traces a symmetric, collimated bipolar jet with the morphology and kinematics that are more typical of protostellar jets. 
In addition, new Gemini adaptive optics images reveal near-IR H$_2$ emission coincident with the H$\alpha$ emission, but not the [Fe~{\sc ii}]. 
Spectra of these three components trace three separate and distinct velocity components: 
(1) H$_2$ from the slow, entrained molecular gas, 
(2) H$\alpha$ from the ionized skin of the accelerating outflow sheath, and 
(3) [Fe~{\sc ii}] from the fast, dense, and collimated protostellar jet itself. 
Together, these data require a driving source inside the dark globule that remains undetected behind a large column density of material. 
In contrast, H$\alpha$ and H$_2$ emission trace the broad outflow of material entrained by the jet, which is irradiated outside the globule. 
As it get dissociated and ionized, it remains visible for only a short time after it is dragged into the H~{\sc ii} region. 
\end{abstract}



\section{Introduction}\label{s:intro}
Protostellar outflows are a beacon of star formation, signaling active disc accretion even in deeply embedded regions where the disc, and sometimes protostars themselves, cannot be detected directly. 
We discuss one such example in this paper. 
The detailed physics of jet launch and collimation is not yet understood \citep[see, e.g.][]{fer06}, but disc accretion ultimately must fuel the jet. 
Observations of relatively unobscured young stars measure this jet directly, but trace an epoch long after the most active accretion. 
CO observations trace outflows from young sources and those in more embedded regions. 
The observed emission may be dominated by ambient molecular material entrained by the jet \citep[e.g.][]{arc05} and/or from a slower disc wind \citep[e.g.][]{kla13}.

Spatially resolved outflows can be a powerful tool to study star formation at higher stellar mass where evidence for discs remains elusive. 
There have been only a few direct detections of disks around intermediate-mass protostars \citep[e.g.][]{kra10,pre11,car12}. 
Structure in the form of clumps and knots along the jet axis points to the variable nature of accretion and outflow, and provides one of the few ways to infer a given protostar's accretion history. 
The longest jets extend parsecs on the sky, sampling a significant fraction of the accretion age \citep[e.g.][]{mar93,dev97,smi04}. 
Many outflows from moderate- to high-mass young stars appear to be scaled-up versions of low-mass systems \citep[e.g.][]{guz11,rei13,rei14}, although this is not always the case \citep[see, e.g.][]{she03}. 
Other outflows, such as the Becklin-Neugebauer / Kleinmann-Low (BN/KL) outflow in Orion OMC1, show evidence for violent processes related to the dynamical interaction and ejection of high-velocity stars that creates an explosive morphology \citep{bal11,god11}. 
However, similar outflow behavior from a large number of protostars over a wide mass range provides compelling evidence that massive star formation can be understood as a scaled-up version of low-mass star formation. 
It is then essential to study intermediate-mass stars ($\sim 2-8$ M$_{\odot}$) where any changes in the dominant physics of formation between low- and high-masses are expected to occur. 

Outflows from young, intermediate-mass protostars are typically observed at millimeter wavelengths where emission from entrained molecules penetrates the high column densities that characterize massive star-forming regions \citep[e.g.][]{tak07,beu08,bel08}. 
However, in an H~{\sc ii} region, much of the obscuring gas and dust may have been cleared. 
In such environments, ultraviolet (UV) radiation from nearby massive stars will illuminate protostellar jets propagating into the H~{\sc ii} region cavity. 
Unlike outflows in quiescent regions where emission at visual wavelengths traces only shock-heated material, external irradiation in H~{\sc ii} regions lights up the body of the entire jet, revealing outflow material that would otherwise remain invisible. 
This allows the physical properties of irradiated jets to be measured using the diagnostics of photoionized gas \citep[e.g.][]{bal01,yus05} with the high angular resolution observations that are available at shorter wavelengths. 

HH~900 is one of the many HH jets discovered by \citet{smi10} as part of their H$\alpha$ survey of the Carina nebula using the Advanced Camera for Surveys (ACS) onboard the \textit{Hubble Space Telescope (HST)}.  
HH~900 is an unusual bipolar outflow emerging from a small tadpole-shaped globule located $\sim 3$ pc (in projection) away from $\eta$ Carinae (see Figure~\ref{fig:hst_ims}). 
In earlier ground-based images, \citet{smi03} identified it as a candidate proplyd, although neither of the two `tails' emerging from the globule point away from $\eta$ Car. 
Higher resolution \textit{HST} images show that these two `tails' appear to be a wide bipolar outflow, and reveal a number of other peculiarities. 
Along the western limb of the broad bipolar outflow lies a strong H$\alpha$ filament and a point source nearly at its center, raising the possibility that the filament is a microjet driven by the star \citep{smi10}. 
Without kinematic information, it remains unclear whether the point source actually drives the putative H$\alpha$ microjet, and if so, whether it is physically related to the larger bipolar outflow. 
Estimated mass-loss rates (derived using the H$\alpha$ emission measure) from the wide inner jet in the eastern limb of the outflow and the microjet along the western limb are the highest of all the outflows found by \citet{smi10}. 

\citet{shi13} present ground-based observations of [Fe~{\sc ii}] 1.64 \micron\ emission in the Carina nebula.  
Based on the morphology of the [Fe~{\sc ii}] emission from HH~900, they propose that the HH~900 driving source is one of the young stellar objects (YSOs) modeled in the Pan-Carina YSO Catalog \citep[PCYC,][]{pov11}. 
A second \textit{Spitzer}-identified YSO lies along the western limb of the flow, although with angular resolution $\gtrsim 1.5$\arcsec, it is unclear how this source relates to the putative microjet. 
\citet{ohl12} conclude that \textit{Herschel} emission near HH~900 is from the externally heated globule and is unlikely to come from young stellar objects in the region.

In this paper, we present new optical and IR observations of HH~900 that allow us to investigate its morphology and kinematics in detail. 
New narrowband IR images obtained with \textit{HST} probe the [Fe~{\sc ii}] 1.26 \micron\ and 1.64 \micron\ lines that are often assumed to be shock-excited in protostellar jets, although this is not necessarily the case in regions with significant FUV radiation. 
Indeed, \citet{rei13} showed that these two [Fe~{\sc ii}] lines probe dense, low-ionization jet material not traced by H$\alpha$. 
Near-IR [Fe~{\sc ii}] lines can also penetrate the extinction inside the dusty birthplaces of these jets, allowing us to connect the larger H$\alpha$ outflows to the \textit{Spitzer}-identified protostars that drive them \citep{rei13}. 
Both [Fe~{\sc ii}] 1.26 \micron\ and 1.64 \micron\ originate from the same $a^4D$ level, so their flux ratio is insensitive to excitation conditions. 
Variations in the flux ratio along the jet therefore provide one way to measure variations in the reddening of the immediate jet environment. 
This may be particularly interesting in HH~900 where extended H$_2$ emission (Hartigan et al. 2015, in preparation) and dark filaments in H$\alpha$ images suggest the presence of molecules and dust in the broad bipolar flow. 

We also present new, second-epoch H$\alpha$ images from \textit{HST} that allow us to measure jet motions in the plane of the sky. 
The outward motion of jet knots provides a direct kinematic identification of candidate jet driving sources. 
Together with ground-based optical and IR spectra, our H$\alpha$ proper motions and [Fe~{\sc ii}] images from \textit{HST} offer a new view of HH~900 that simultaneously help to unravel but also deepen the mystery of this unusual outflow.


\section{Observations}\label{s:obs}
\begin{table*}
\caption{Observations\label{t:obs}}
\centering
\begin{tabular}{lllll}
\hline\hline
Instrument & Filter / Position & Date & Int. time & Comment \\
\hline
ACS/\textit{HST} & F658N & 2005 Jul 18 & 1000s & H$\alpha$ $+$ [N~{\sc ii}] \\
\vspace{3pt}
ACS/\textit{HST} & F658N & 2014 Aug 04 & 1000s & H$\alpha$ $+$ [N~{\sc ii}] \\
GSAOI/\textit{Gemini} & K$_\mathrm{s}$ N cont. & 2013 Mar 23 & 1080s & \\
\vspace{3pt}
GSAOI/\textit{Gemini} & H$_2$ & 2013 Mar 23 & 1080s & \\
WFC3-IR/\textit{HST} & F126N & 2013 Dec 28 & 2397 s & [Fe~{\sc ii}] $\lambda12567$ \\
WFC3-IR/\textit{HST} & F128N & 2013 Dec 28 & 2397 s & continuum \\
WFC3-IR/\textit{HST} & F164N & 2013 Dec 28 & 2397 s & [Fe~{\sc ii}] $\lambda16435$ \\
\vspace{3pt}
WFC3-IR/\textit{HST} & F170N & 2013 Dec 28 & 2397 s & continuum \\
\vspace{3pt}
EMMI/\textit{NTT} & H$\alpha$, [S~{\sc ii}] & 2003 Mar 09 & 900 s & P.A. $=63^{\circ}$ \\
FIRE/\textit{Magellan} & West & 2014 Jan 15 & 600 s & P.A. $=77^{\circ}$ \\
FIRE/\textit{Magellan} & East & 2014 Jan 16 & 600 s & P.A. $=60^{\circ}$ \\
\hline
\end{tabular}
\label{t:foobar}
\end{table*}
\subsection{High-resolution images} 
\textbf{[Fe~{\sc ii}]:}
Table~\ref{t:obs} lists the details of the images and spectroscopy presented in this paper. 
Near-IR, narrowband [Fe~{\sc ii}] images and corresponding off-line narrowband continuum images were obtained with WFC3-IR on board the \textit{Hubble Space Telescope (HST)} under program GO-13391 (PI: N. Smith) in Cycle 21. 
For HH~900, we duplicated the observing strategy used previously to obtain [Fe~{\sc ii}] images of four other jets in the Carina nebula, HH~666, HH~901, HH~902, and HH~1066 \citep{rei13}. 
We employed a box-dither pattern to avoid dead-pixel artifacts and to provide modest resolution enhancement. 
Using the same integration time for all filters provided a similar signal-to-noise ratio for the [Fe~{\sc ii}] and adjacent continuum images.

\textbf{H$_2$:}
We obtained near-IR, narrowband H$_2$ 2.12 \micron\ images and complementary off-line narrowband K(short) continuum images (central wavelength of 2.093 \micron) during early science with the Gemini South Adaptive Optics Imager \citep[GSAOI,][]{mcg04,gem_car12}. 
GSAOI is a near-IR imager, used in combination with Gemini Multi-Conjugate Adaptive Optics System \citep[GeMS,][]{rig14,nei14} with natural guide stars to provide near-diffraction limited images over a field of view of 85\arcmin\ $\times$ 85\arcmin\ in the 0.9-2.5 \micron\ range. 
HH~900 was observed 22 March 2013 in queue observing mode with a 9-point dither pattern and individual integrations of 120 s. 
We resampled the aligned, flat-fielded images to the same pixel scale as the WFC3-IR images. 
Simultaneous K(short) continuum images allow us to subtract bright IR continuum emission from the edge of the globule and therefore search for extended H$_2$ emission from the jet. 

\textbf{H$\alpha$:} 
H$\alpha$ images of HH~900 were obtained with \textit{HST}/ACS for the first time on 18 July 2005 \citep{smi10}. 
We obtained a second epoch on 04 August 2014 using the the same instrument and filter, duplicating the first-epoch observational setup in order to measure proper motions (program GO-13390, PI: N. Smith). 
Together, this provides a $\sim 9$ yr time baseline between observations, allowing us to measure the motion of faint jet features to $\sim 25$ km s$^{-1}$. 
Using the same orientation and coordinates for both epochs minimizes position-dependent errors when determining proper motions.

We follow a method similar to that of \citet{and08a,and08b}, \citet{and10}, and \citet{soh12} to align and then measure precise proper motions in the ACS images. 
This method is based on PSF photometry of the bias-subtracted, flat-fielded, and CTE-corrected {\tt flc} images produced by the \emph{HST} pipeline.  
Unlike the drizzled {\tt drc} images, which have been calibrated, flat-fielded, CTE- and geometrically-corrected, and dither-combined (via AstroDrizzle), the {\tt flc} images have not been resampled to correct for geometric distortion and thus allow for more accurate PSF fitting.

In brief, we first measure centroid positions for bright, relatively isolated stars in the {\tt drc} images, since these images have an astrometric solution that allows us to construct an initial reference frame. 
As in \citet{and08a}, our reference frame has a 50 mas pixel$^{-1}$ scale and a north-aligned \emph{y} axis. 
Next, we perform PSF photometry on the individual flc exposures using the program {\tt img2xym\_WFC.09x10} \citep{and06}, which uses a library of models of the spatially-variable effective ACS PSF. 
Once measured, stellar positions are corrected for distortion using the \citet{and05} corrections. 
Stars in common between the distortion-corrected frame of each {\tt flc} exposure and the reference frame are identified, allowing us to determine six-parameter linear transformations from the former to the latter.  We iterate on this process once, replacing the centroid positions that initially defined the reference frame with the average transformed reference-frame positions from the {\tt flc} PSF photometry, then recomputing the linear transformation from each {\tt flc} image to the reference frame. 

We relate both epochs to a single reference frame, so the measured positions in each epoch can be directly compared. Relative proper motions for point sources (including the YSOs adjacent to HH~900) are simply the difference in average reference frame position between 2005 and 2014. The positional errors in each dimension for each epoch are calculated by dividing the rms scatter of the individual measurements from each exposure, divided by the square root of the number of exposures (N=2-4) used to compute the average position.  

Our \emph{HST} observations are such that the western bow shock of HH~900 was observed in a different orbit than the globule, inner jet, and eastern bow shock. 
Each orbit consists of three overlapping pointings, with two dithered exposures per pointing; however, the overlap in area between orbits is small. 
This requires that we construct a separate reference frame for each orbit. 
These reference frames are defined by the positions of the stars and are not tied to an absolute proper-motion zero point. 
Therefore, all motions are measured relative to the average motion of the stars in the image.  
Because the HH~900 western bow shock was observed in a separate frame, there may be a systematic offset between the proper motion of the western bow shock and the rest of HH~900. We expect this offset to be $\lesssim1$ km s$^{-1}$, but certainly no larger than the stellar velocity dispersion of the Tr 16 cluster, or $\lesssim5$ km s$^{-1}$.  

To measure the proper motions of extended features in HH~900, we create stacked images by resampling the {\tt flc} exposures into our reference frame \citep[see][for details of the stacking algorithm used]{and08a}.
We then select bright jet features that do not change significantly in morphology between the two epochs and measure their motion using the modified cross-correlation technique described by \citet{cur96,har01,mor01}. After subtracting a median-filtered image to account for background emission, we select a small box around each jet feature. 
This subimage is shifted relative to the reference image, and for each shift, we compute the total of the square of the difference over the box region between the two images. The minimum of this array corresponds to the pixel offset between the two images. 
For the typical signal-to-noise in our images, the uncertainty of this procedure is $\approx 0.03$ pixel, corresponding to $\approx 10$ km s$^{-1}$.  

We also measure the proper motions of point sources in the image by fitting a spatially variable PSF to the stars. 
This allows the motion of stars to be measured to greater precision than nebulous jet features, typically a few km s$^{-1}$ (see Table~2).

\begin{figure}
\centering
$\begin{array}{c}
\includegraphics[trim=10mm 0mm 0mm 0mm,angle=0,scale=0.45]{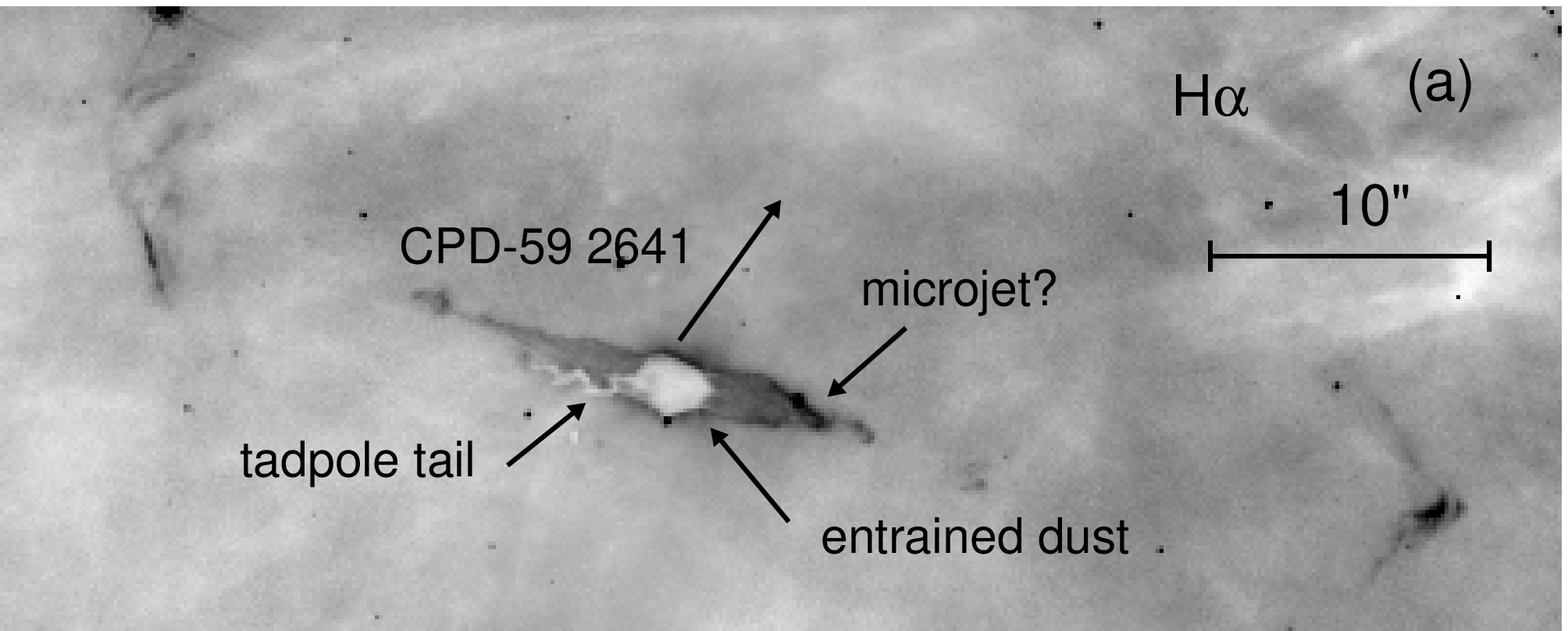} \\ 
\includegraphics[trim=10mm 0mm 0mm 0mm,angle=0,scale=0.45]{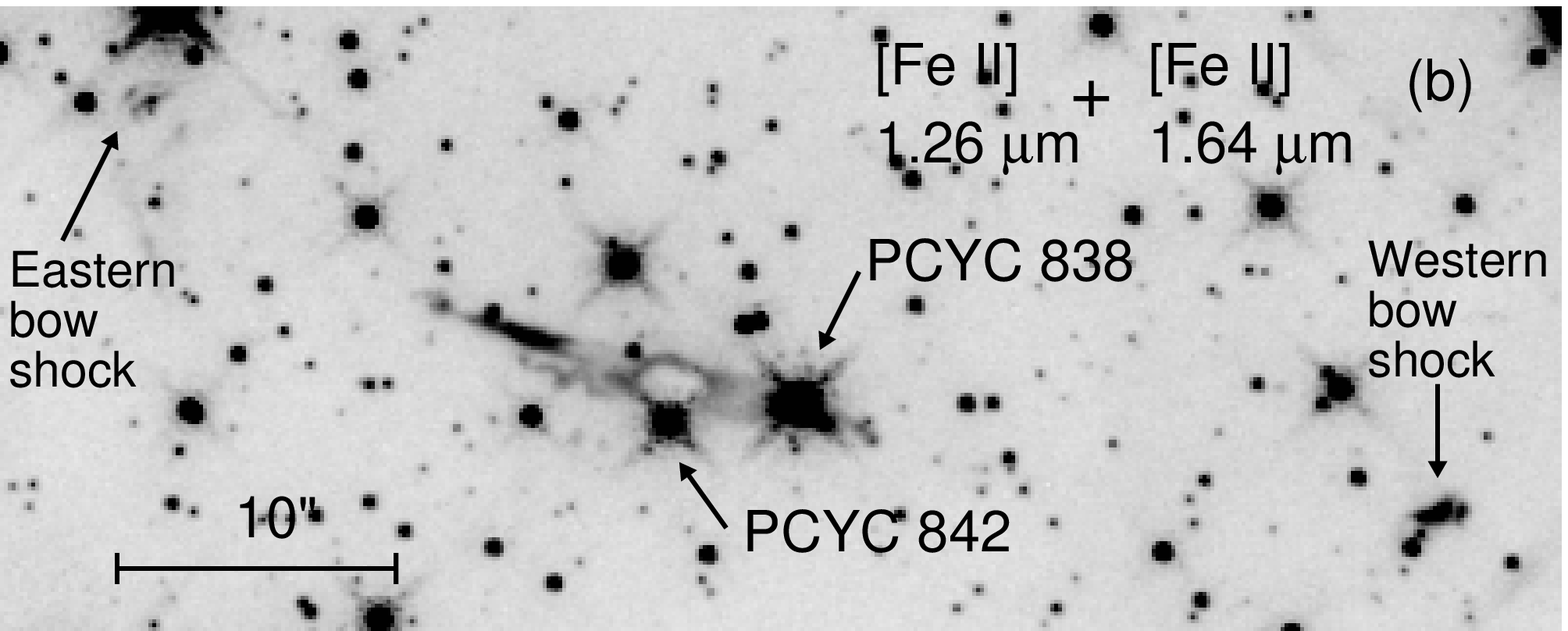} \\
\includegraphics[trim=10mm 0mm 0mm 0mm,angle=0,scale=0.45]{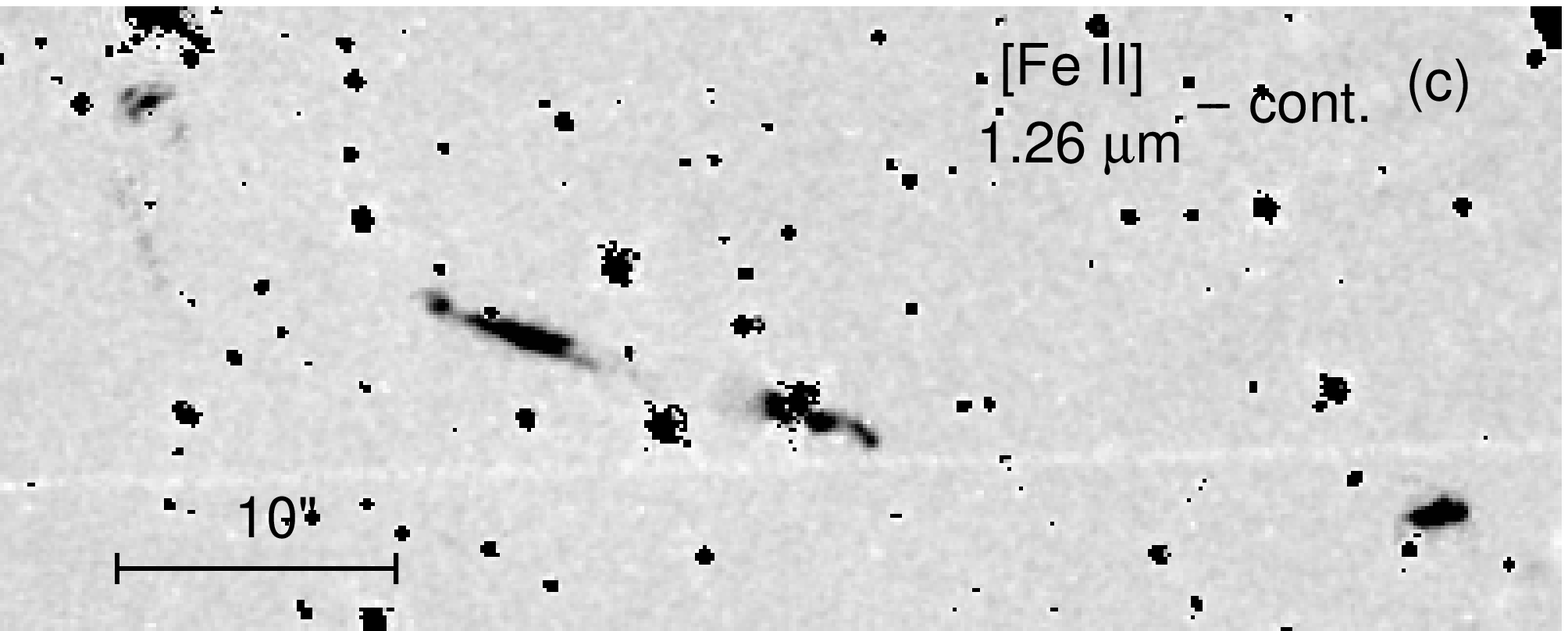} \\ 
\includegraphics[trim=10mm 0mm 0mm 0mm,angle=0,scale=0.45]{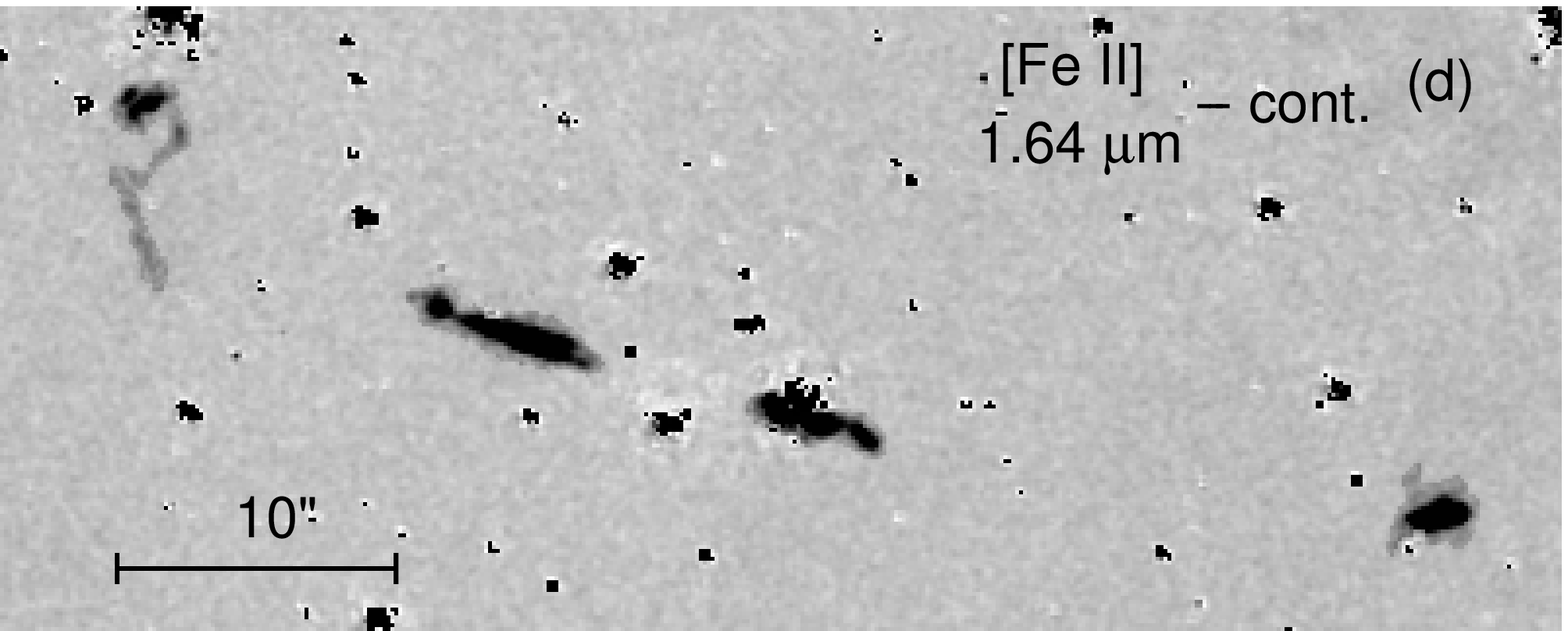} \\
\end{array}$
\caption{New narrowband [Fe~{\sc ii}] images of HH~900 obtained with \textit{HST}/WFC3-IR. 
Panel (a) is the \textit{HST}/ACS H$\alpha$ image from \citet{smi10} showing the full extent of the optical outflow. 
Panel (b) is a [Fe~{\sc ii}] 1.26 \micron\ $+$ [Fe~{\sc ii}] 1.64 \micron\ image of HH~900 with the two YSOs identified by \citet{shi13} from the PCYC labelled.  
Panels (c) and (d) show the continuum-subtracted [Fe~{\sc ii}] 1.26 \micron\ and 
[Fe~{\sc ii}] 1.64 \micron\ images, respectively. 
 }\label{fig:hst_ims} 
\end{figure}

\subsection{Spectroscopy}
\textbf{FIRE:}
We obtained near-IR spectra of HH~900 with the Folded-Port InfraRed Echellette (FIRE) near-IR spectrograph \citep{sim08,sim10,sim13} on the Magellan Baade 6.5-m telescope on 17 January 2014 (see Table~\ref{t:obs}). 
FIRE's $0.8 - 2.5$ \micron\ wavelength coverage includes multiple emission lines from the jet. 
We positioned the 7\arcsec\ slit along bright [Fe~{\sc ii}] emission in the eastern and western limbs of the outflow and used a 1\arcsec\ slit width to accommodate $\gtrsim 1\farcs1$ seeing. 
Figure~\ref{fig:feii_pvs} shows the slit positions. 
The YSO and putative microjet that lie along the western limb of the larger bipolar outflow fall within the slit used to observe the western limb of the [Fe~{\sc ii}] jet. 
To account for sky emission, we employed a nodding strategy, pointing on- and off-source in an ABBA sequence with an average sky offset located $\sim2$\arcsec\ below the jet. 
Wavelength calibration was done using the internal ThAr lamp and data reduction was performed with the {\sc Firehose IDL} pipeline.  
After correcting for the motion of the earth, we report Doppler velocities in the heliocentric velocity frame. 

\textbf{EMMI:}
We obtained visual-wavelength spectra of HH~900 with the ESO Multi-Mode
Instrument (EMMI) on the New Technology Telescope (NTT) on 09 March 2003. 
We observed HH~900 in echelle mode, using the cross-dispersing grism to obtain 
H$\alpha$ and [S~{\sc ii}] simultaneously. 
The echelle slit length is 25\arcsec\ and the jet was observed with a 1\arcsec\ 
slit width. 
We aligned the slit through the middle of the globule and the star in the western 
limb of the flow. 
We estimate sky emission from separate frames, with the slit offset 
$\sim 4$\arcsec\ perpendicular to the jet. 
We compute the wavelength solution from the sky frames, using the nebular [N~{\sc ii}] 
lines for the H$\alpha$ dispersion solution and the [S~{\sc ii}] lines themselves. 
Thus our derived radial velocities are relative to the Doppler shift of ambient ionized gas in the Carina H~{\sc ii} region. 


\section{Results}\label{s:results}
\begin{figure}
\centering
$\begin{array}{c}
\includegraphics[trim=10mm 0mm 0mm 0mm,angle=0,scale=0.55]{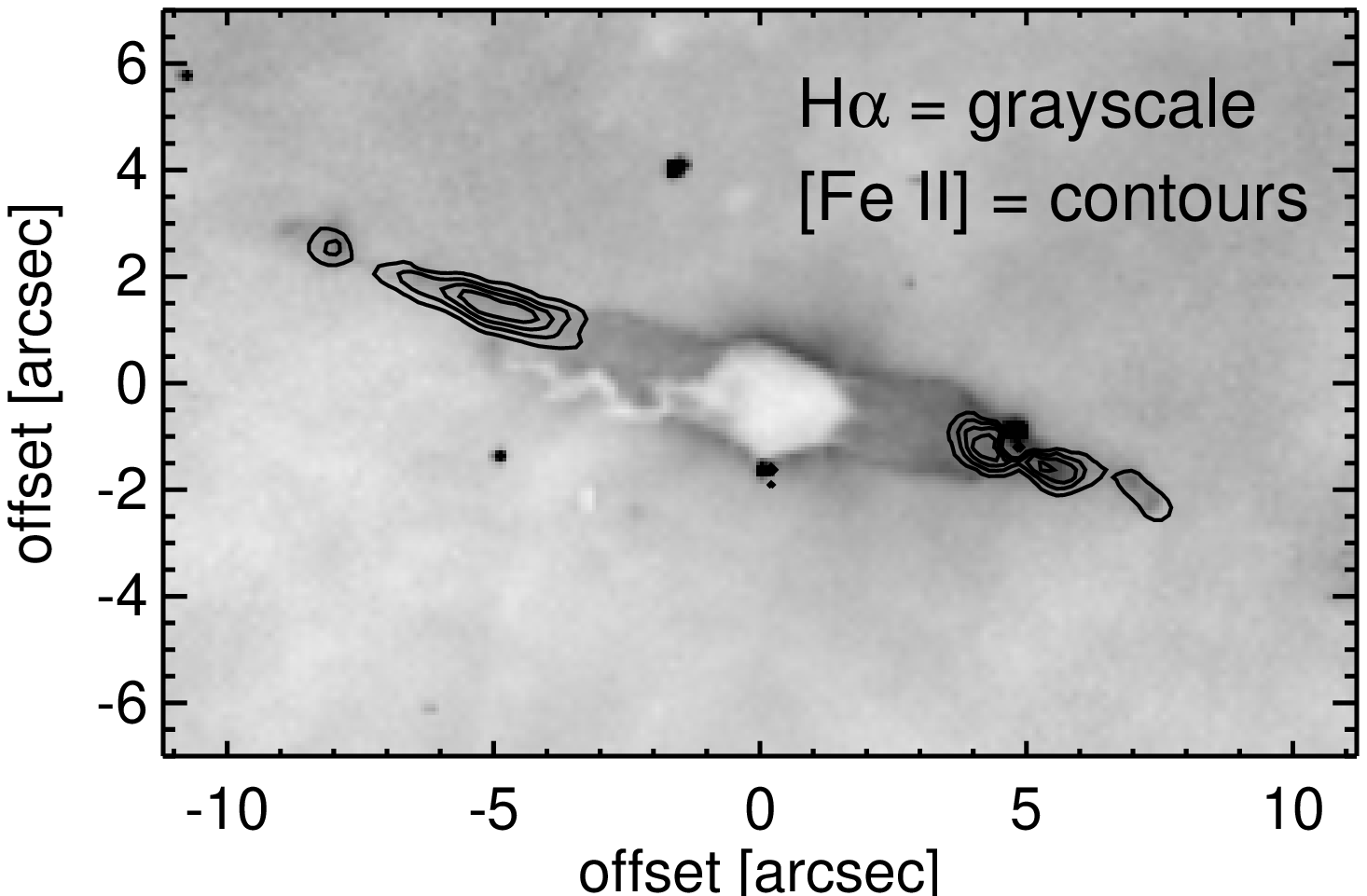} \\ 
\includegraphics[trim=10mm 0mm 0mm 0mm,angle=0,scale=0.55]{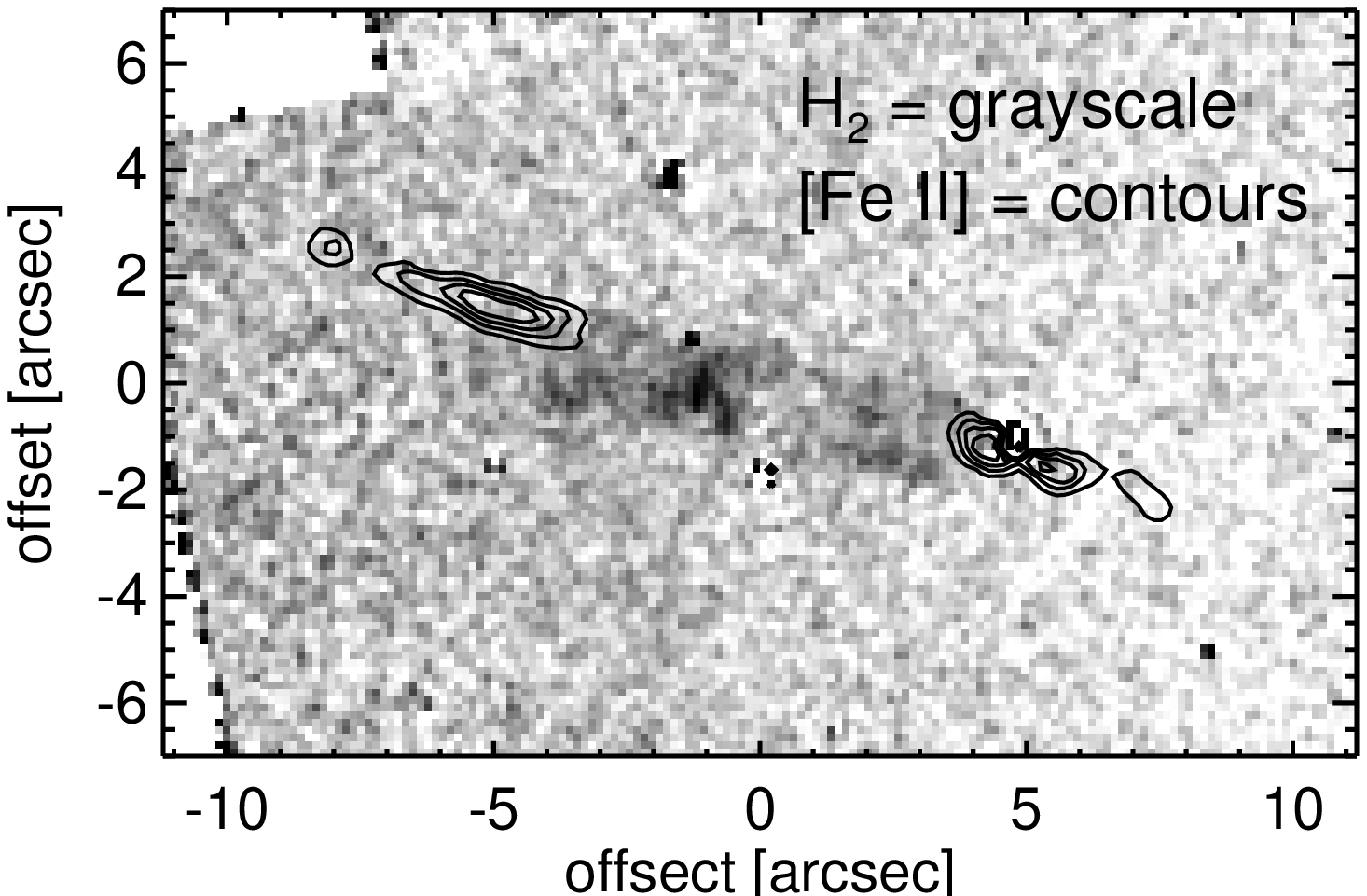} \\ 
\end{array}$
\caption{H$\alpha$ \textit{HST}/ACS image from \citet{smi10} (top) and narrowband H$_2$ $-$ K-band image from \textit{GSAOI} (bottom), both with [Fe~{\sc ii}] contours (1.26 \micron\ $+$ 1.64 \micron) overplotted. [Fe~{\sc ii}] emission begins at the same place that H$_2$ emission fades below our sensitivity and the H$\alpha$ emission narrows to a similar width as the [Fe~{\sc ii}] emission. 
}\label{fig:fe_contours} 
\end{figure}
\begin{figure}
\centering
$\begin{array}{c}
\includegraphics[trim=-5mm 0mm 0mm 0mm,angle=0,scale=0.375]{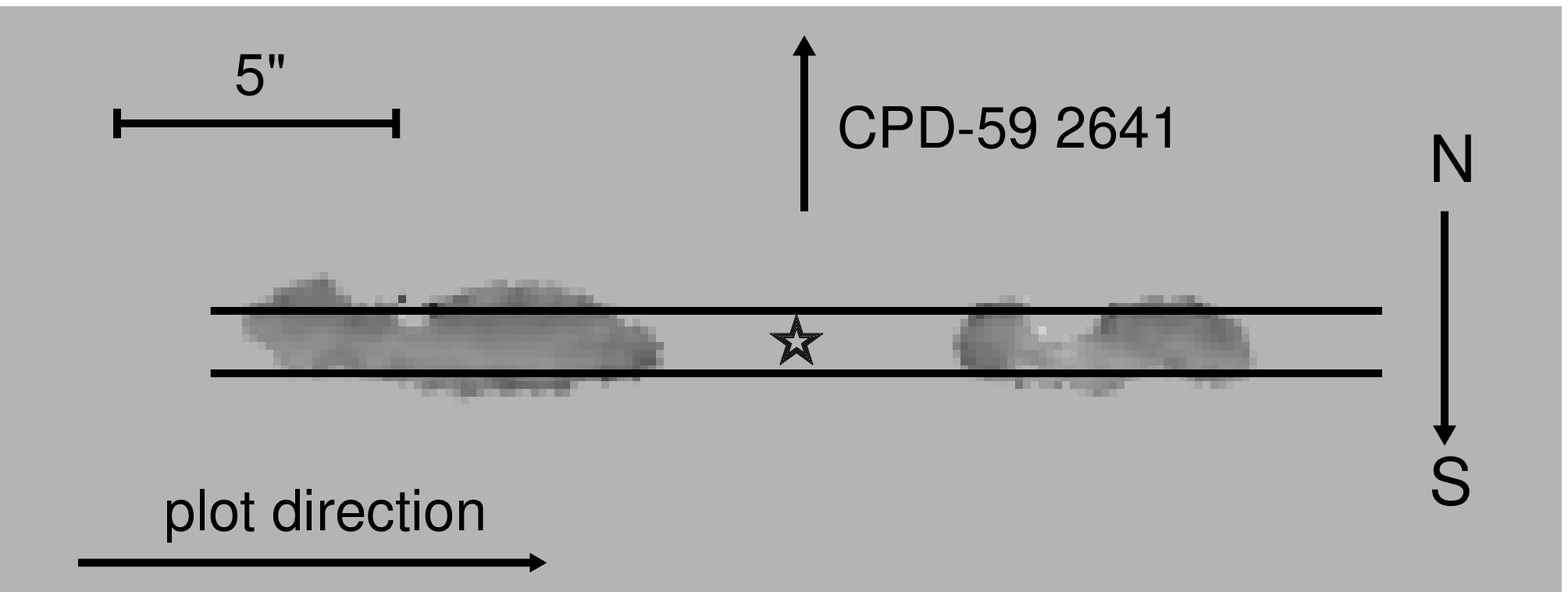} \\ 
\includegraphics[angle=90,scale=0.325]{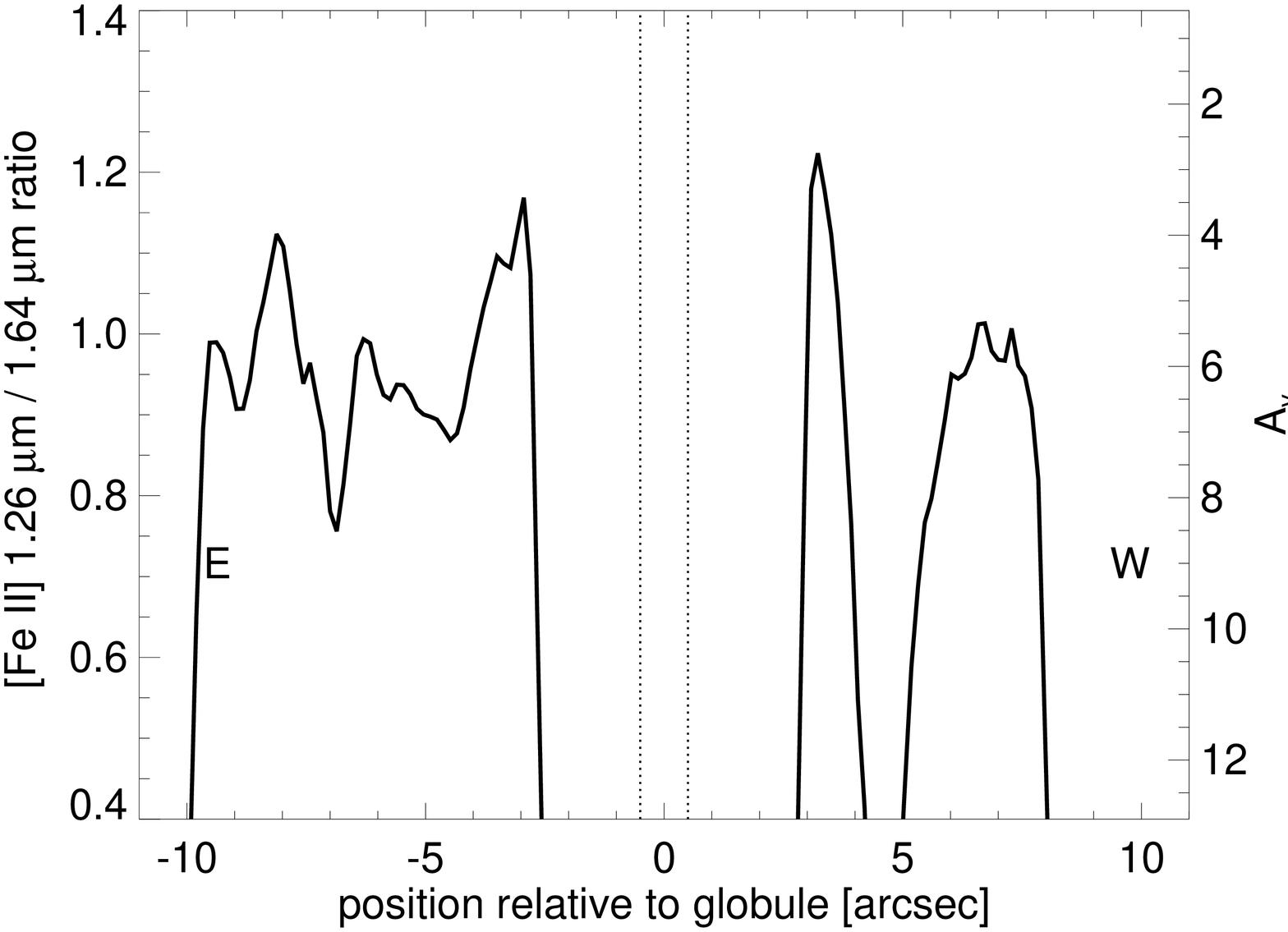} \\ 
\includegraphics[angle=90,scale=0.325]{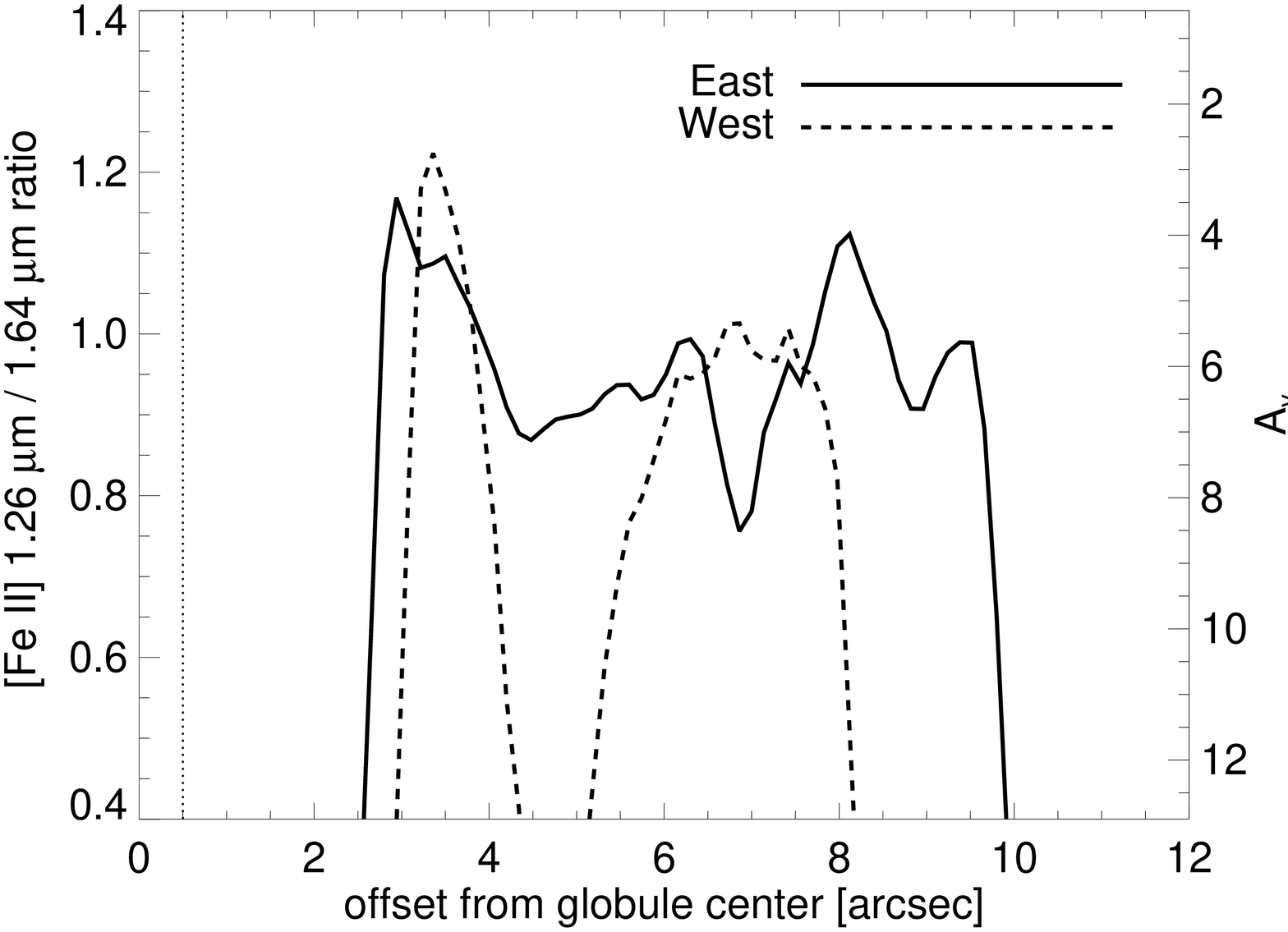} \\ 
\end{array}$
\caption{
\textit{Top:} A [Fe~{\sc ii}] 1.26 \micron/ 1.64 \micron\ ratio image (darker colours correspond to a higher value of $\mathcal{R}$) with lines indicating the boundaries used to make the [Fe~{\sc ii}] ratio tracings along the length of the jet below plotted below. 
\textit{Middle:} Tracing of the inner jet along the [Fe~{\sc ii}] jet axis with positions plotted as a function of the offset from the approximate location of the center of the globule. 
\textit{Bottom:} Same tracing as above, but with the two sides of the jet overplotted on top of each other. 
}\label{fig:feii_tracings} 
\end{figure}
\subsection{Morphology in the IR images}\label{ss:morph}
\begin{figure}
\centering
$\begin{array}{c}
\includegraphics[angle=0,scale=0.375]{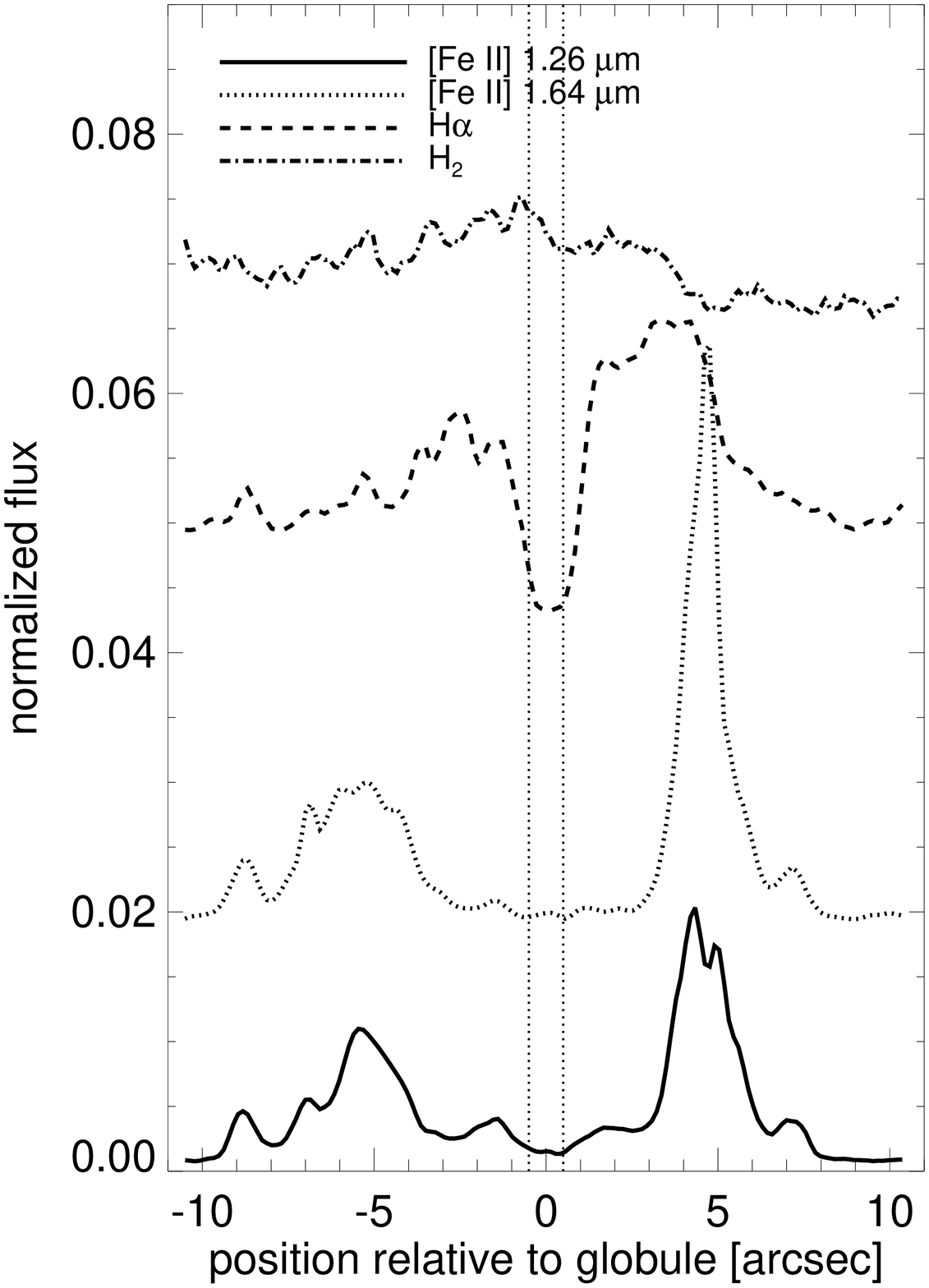} \\
\includegraphics[angle=0,scale=0.375]{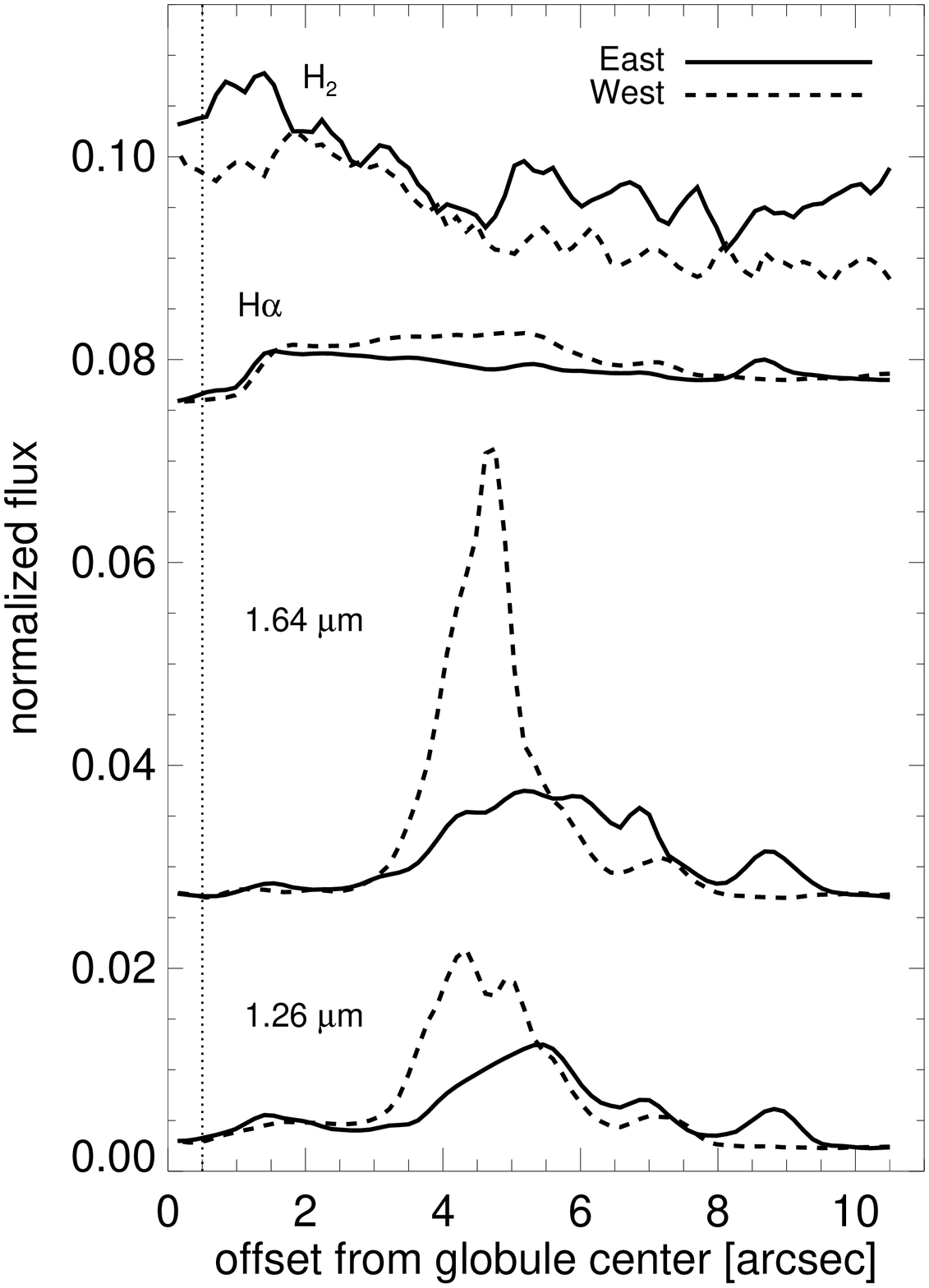} \\
\end{array}$
\caption{\textit{Top:} Median intensity traced along the axis of the [Fe~{\sc ii}] jet in HH~900 (using the same tracing area shown in Figure~\ref{fig:feii_tracings}). 
\textit{Bottom:} Median intensity in the eastern and western limbs of the jet overplotted on each other. 
}\label{fig:feii_ratio_tracings} 
\end{figure}

Figure~\ref{fig:hst_ims} shows the new [Fe~{\sc ii}] 1.26 \micron\ and 1.64 \micron\ images of HH~900 obtained with \textit{HST}/WFC3-IR (Figures~\ref{fig:hst_ims}b, c, and d) and the ACS H$\alpha$ image (Figure~\ref{fig:hst_ims}a). 
Bright, collimated [Fe~{\sc ii}] emission that extends to the east and west of the dark tadpole globule stands out in the continuum-subtracted images (Figures~\ref{fig:hst_ims}c and d). 
Near-IR [Fe~{\sc ii}] emission traces a symmetric, collimated bipolar jet, unlike H$\alpha$, which emerges into the H~{\sc ii} region with almost the same geometric width as the globule (see Figure~\ref{fig:fe_contours}). 
Figure~\ref{fig:fe_contours} shows a more detailed comparison of the H$\alpha$ and [Fe~{\sc ii}]. 
Intensity tracings through the eastern and western limbs of the [Fe~{\sc ii}] jet show that they are remarkably similar (see Figure~\ref{fig:feii_tracings}~and~\ref{fig:feii_ratio_tracings}).

Both sides of the jet show a $\gtrsim 1.5$\arcsec\ gap between the edge of the globule and bright [Fe~{\sc ii}] emission from the jet. 
Unlike the images of jets in our previous study using archival data \citep{rei13}, our observations of HH~900 include simultaneous line-free continuum images that allow us to subtract the continuum and therefore separate photodissociation region (PDR) emission along the globule edge from fainter [Fe~{\sc ii}] emission emitted by the inner jet. 
Subtracting the line-free continuum from HH~900 demonstrates that the offset of the [Fe~{\sc ii}] jet from the edge of the globule is real (see Figure~\ref{fig:fe_contours}). 
No tenuous [Fe~{\sc ii}] emission from the inner jet extends back to the globule edge in the continuum-subtracted image, demonstrating that the offset is not just confusion with continuum emission that grows increasingly bright close to the irradiated edge of the dusty globule. 

%
\begin{figure}
\centering
$\begin{array}{c}
\includegraphics[trim=0mm 0mm 0mm 0mm,angle=0,scale=0.40]{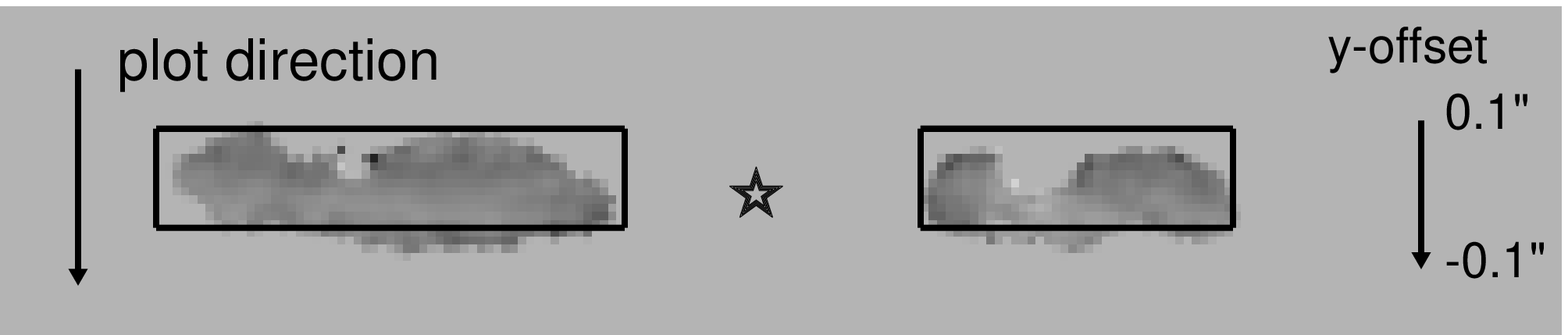} \\
\begin{array}{cc}
\includegraphics[trim=30mm 0mm 10mm 0mm,angle=0,scale=0.25]{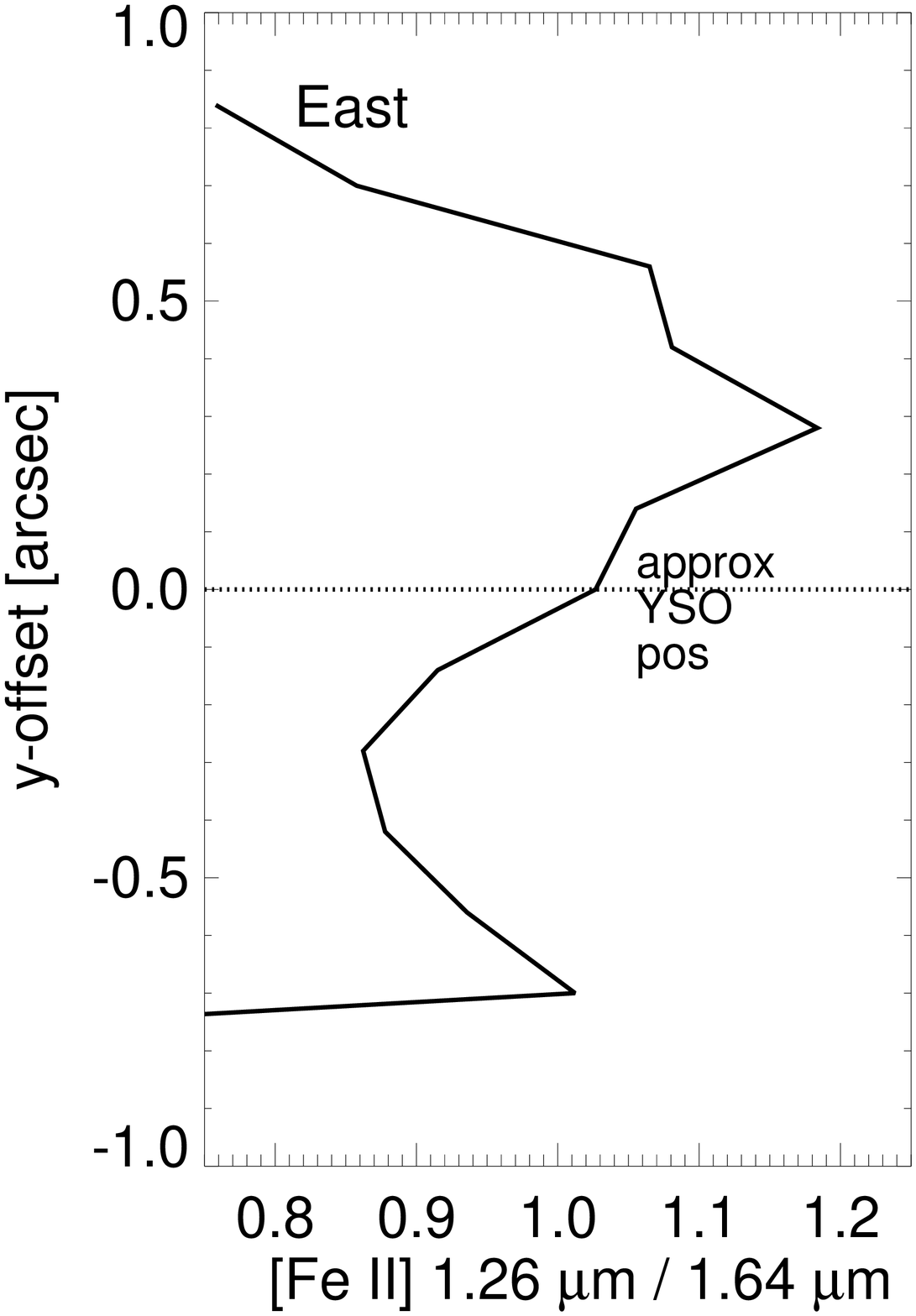} &
\includegraphics[trim=40mm 0mm 0mm 0mm,angle=0,scale=0.25]{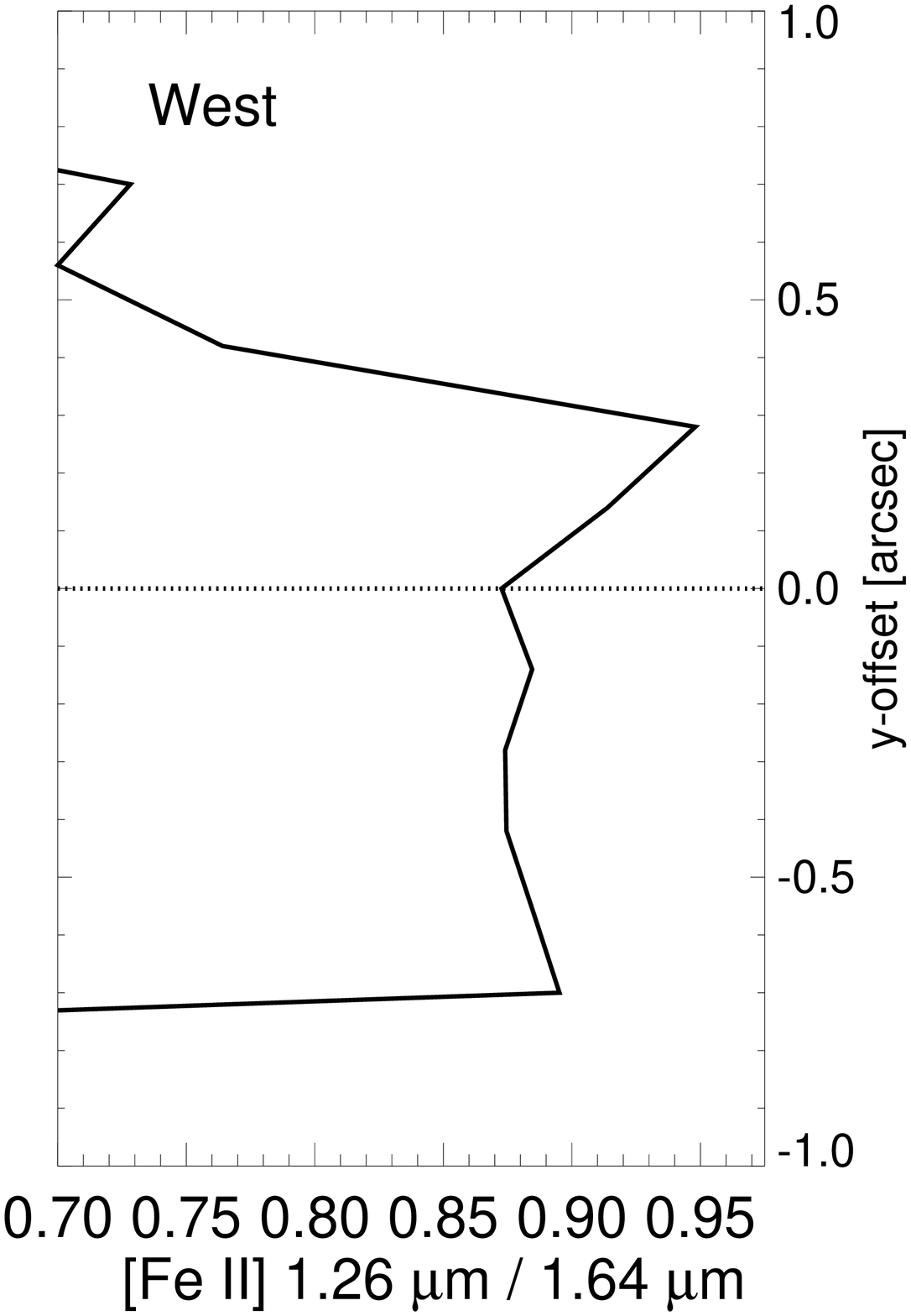} \\
\end{array}
\end{array}$
\caption{\textit{Top:} [Fe~{\sc ii}] 1.26 \micron/ 1.64 \micron\ ratio image with lines showing the boxes used to extract the tracings through the jet. A star in the middle denotes the estimated location of the invisible driving source. 
\textit{Bottom left:} Plots of the average [Fe~{\sc ii}] 1.26 \micron\ / 1.64 \micron\ ratio through the \textit{width} of the eastern limb of the jet (perpendicular to the direction of jet propagation). We sum the emission along the length of the jet, as shown in the image at top. 
Positive offsets on the y-axis are located above the jet, negative offsets fall below. 
\textit{Bottom right:} The average [Fe~{\sc ii}] 1.26 \micron\ / 1.64 \micron\ ratio through the western limb of HH~900. 
}\label{fig:feii_y_tracing} 
\end{figure}

\citet{smi10} identify two bow shocks associated with HH~900 that are $\sim 10$\arcsec\ beyond the end of the inner jet in H$\alpha$ images. 
Both bow shocks are also bright in [Fe~{\sc ii}] emission (see Figure~\ref{fig:hst_ims}).

Three dark, filamentary streaks in the western limb of HH~900 are seen in absorption in the H$\alpha$ image (see Figure~\ref{fig:hst_ims}a). 
\citet{smi10} interpret these dark filaments as extinction due to dust entrained by the jet along the walls of the outflow cavity, similar to the features that \citet{wal04} observed in HH~280 (another jet that appears to burst out of a globule). 
Bright emission extending off the western edge of the globule in the [Fe~{\sc ii}] image in Figure~\ref{fig:hst_ims}a disappears in the continuum-subtracted image, indicating that it is continuum emission. 
This may be starlight scattered by the dusty filaments. 

Narrowband H$_2$ images from \textit{GSAOI} show H$_2$ emission from the surface of the globule and outside it that, like the H$\alpha$ emission, appears broad with nearly the same projected width as the globule itself (see Figure~\ref{fig:fe_contours}). 
The intensity of the H$_2$ emission appears to be slightly ($\sim 1.5\times$) greater along the edges of the outflow than in the center (i.e. limb brightened). 
H$_2$ emission from the outflow fades below our detection limit $\sim 1\farcs5$ away from the globule edge; this is the same place where the width of the H$\alpha$ emission appears to decrease and the collimated [Fe~{\sc ii}] jet begins (see Figure~\ref{fig:fe_contours}).

\subsection{[Fe~{\sc ii}] ratio}\label{ss:ratio}
Both the [Fe~{\sc ii}] 1.26 \micron\ and 1.64 \micron\ lines originate from the same upper level, so the observed flux ratio therefore provides a measure of the reddening. 
In unobscured environments, the 1.26 \micron\ line will be brighter, with a derived flux ratio, $\mathcal{R} = \lambda12567$/$\lambda16435 = 1.49$ \citep{sh06}. 
Smaller values of $\mathcal{R}$ indicate more reddening and extinction. 
Along the length of the [Fe~{\sc ii}] jet, the ratio remains consistent, with a value of $\sim 0.95$ (see Figure~\ref{fig:feii_tracings}). 
This corresponds to an E(B$-$V) $\approx 0.9$ \citep{rei13}. 
Using the ratio of total to selective extinction measured toward Carina of $R=4.8$ \citep{smi87,smi02}, this corresponds to $A_V \approx 4.5$ mag along the length of the jet. 
We also measure $\mathcal{R}$ through the width of the jet (perpendicular to the direction of propagation, see Figure~\ref{fig:feii_y_tracing}). 
Reddening through the width of the jet increases further away from $\eta$ Car and Tr 16, corresponding to a decrease in $\mathcal{R}$ of $\sim 0.2$ across the width of the jet. 
The measured decrease in $\mathcal{R}$ corresponds to $\sim 1.5$ mag more extinction on the side of the jet further away from the ionizing source. 

\subsection{Candidate driving sources}\label{ss:ysos}
\citet{rei13} showed that in HH~666 and HH~1066, the driving protostar lies along the projected jet axis and that [Fe~{\sc ii}] connects the jet to the \textit{Spitzer}-identified driving sources. 
However, [Fe~{\sc ii}] emission in HH~900 does not reach back to the edge of the tadpole-shaped globule, and no [Fe~{\sc ii}] emission from the jet is detected inside the globule. 

Two IR-bright YSOs have been found in the vicinity of the HH~900 globule \citep{shi13}. 
The jet axis traced by [Fe~{\sc ii}] emission lies $\sim 1.5$\arcsec\ north of the YSO that lies at the bottom of the tadpole globule \citep[PCYC 842;][see Figure~\ref{fig:hst_ims}]{pov11}, clearly demonstrating that it cannot be the driving source of the bipolar flow. 
The second \textit{Spitzer}-identified YSO (PCYC~838) lies along the western limb of the flow, in the middle of the putative microjet. 
While the morphology raises the possibility that this protostar drives a separate jet that happens to be aligned with the larger H$\alpha$ outflow lobe, images alone do not indicate whether there is a separate microjet or if PCYC~838 is its driving source. 
As we discuss in Section~\ref{ss:pm}, however, radial velocities and proper motions lead us to reject this hypothesis. 

Evidence for an additional protostellar source along the jet axis embedded in the tadpole globule remains elusive due to the small size of the globule ($\sim 2$\arcsec) and the coarser resolution of available mid- and far-IR observations. 
We do not detect any evidence of an IR-bright point source inside the globule in any of the WFC3-IR images. 
The upper limit in all four bands is $\geq 21$ mag. 
High frequency, high spatial resolution observations with ALMA are required to determine whether there is an additional protostar embedded in the globule.

\subsection{Velocity structure}\label{ss:velocity}
\begin{figure*}
\centering
$\begin{array}{ccc}
\includegraphics[trim=100mm 0mm 0mm 10mm,angle=90,scale=0.60]{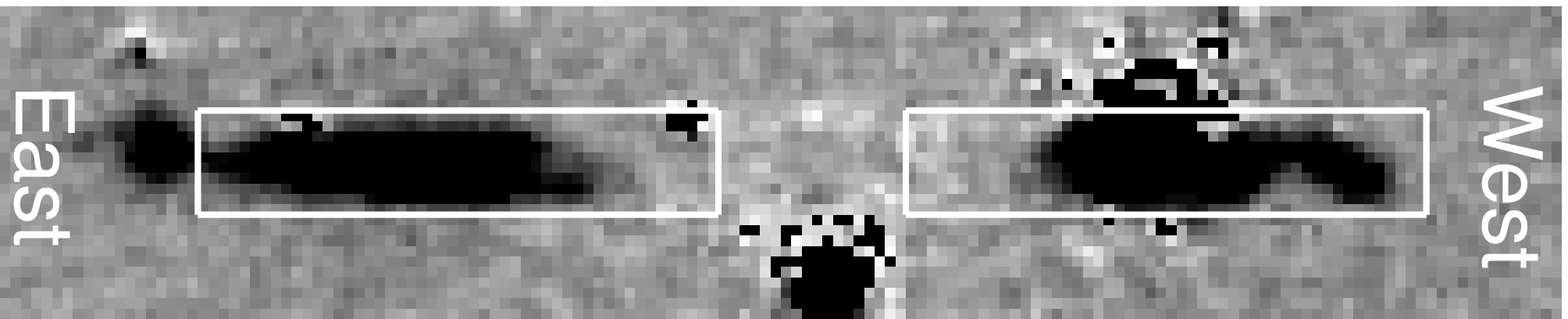} & 

\begin{array}{c}
\includegraphics[trim=0mm 10mm 10mm 0mm,angle=0,scale=0.325]{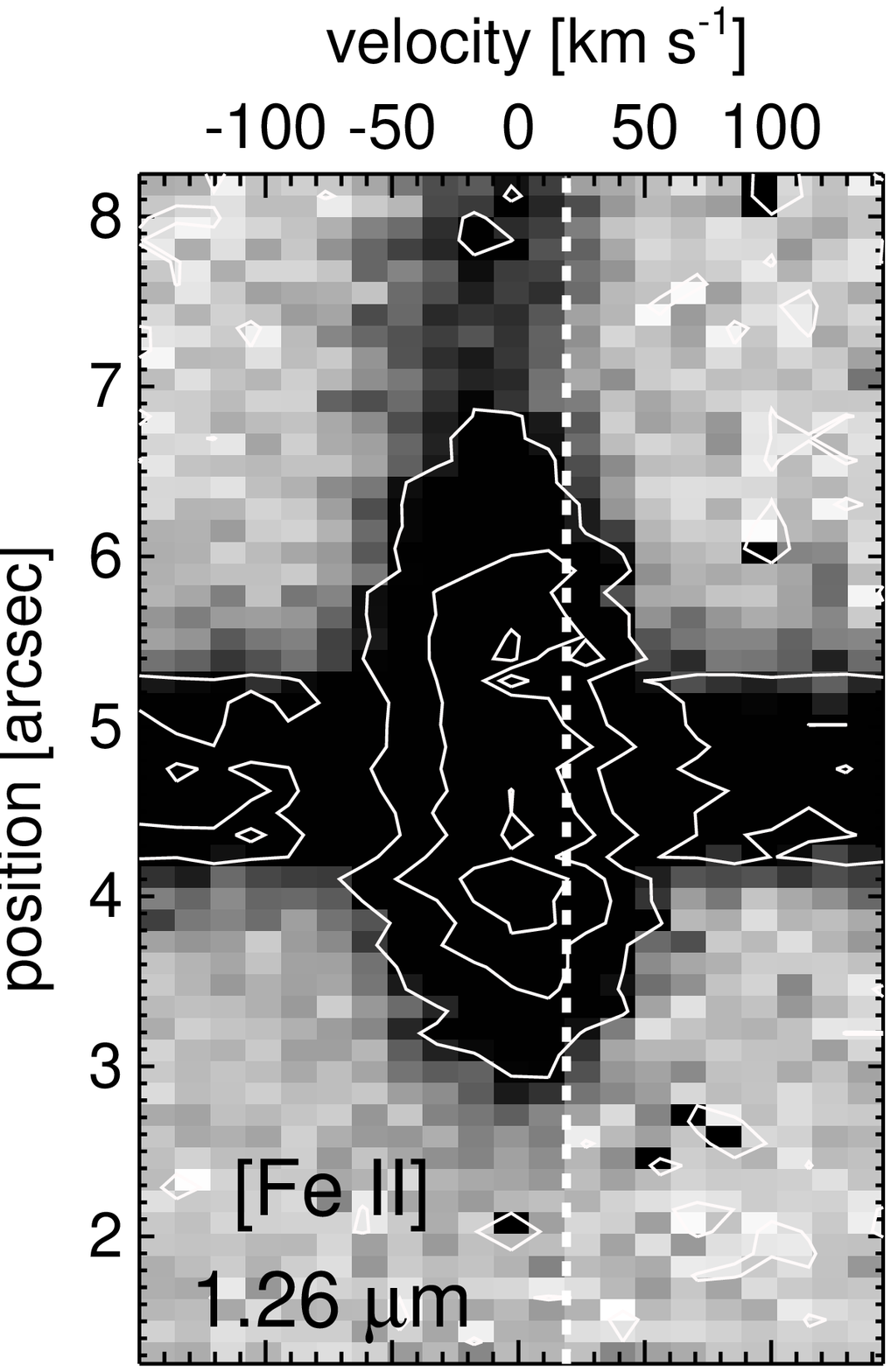} \\
\includegraphics[trim=0mm 0mm 10mm 10mm,angle=0,scale=0.325]{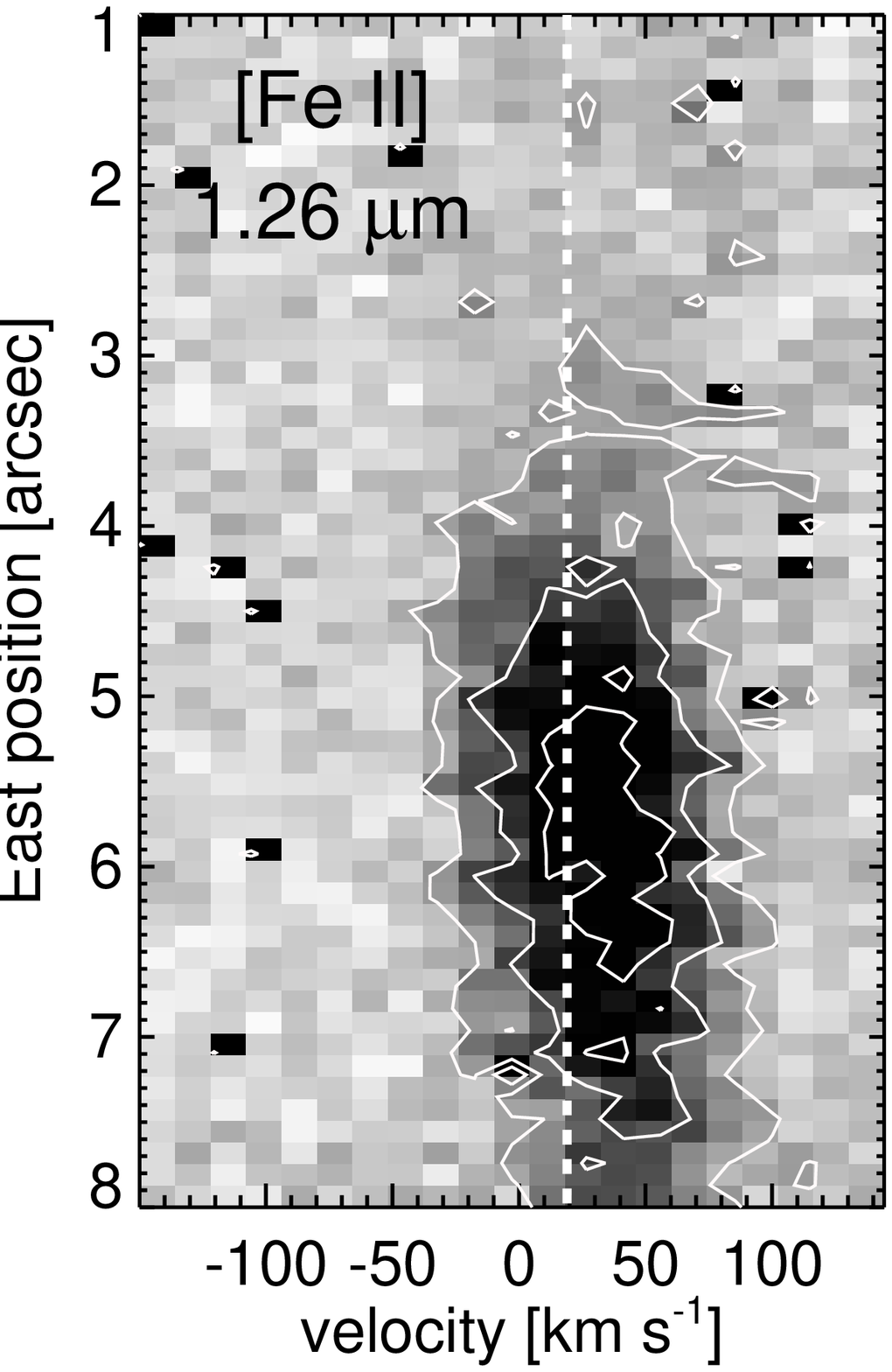} \\
\end{array}

\begin{array}{c}
\includegraphics[trim=20mm 10mm 10mm 0mm,angle=0,scale=0.325]{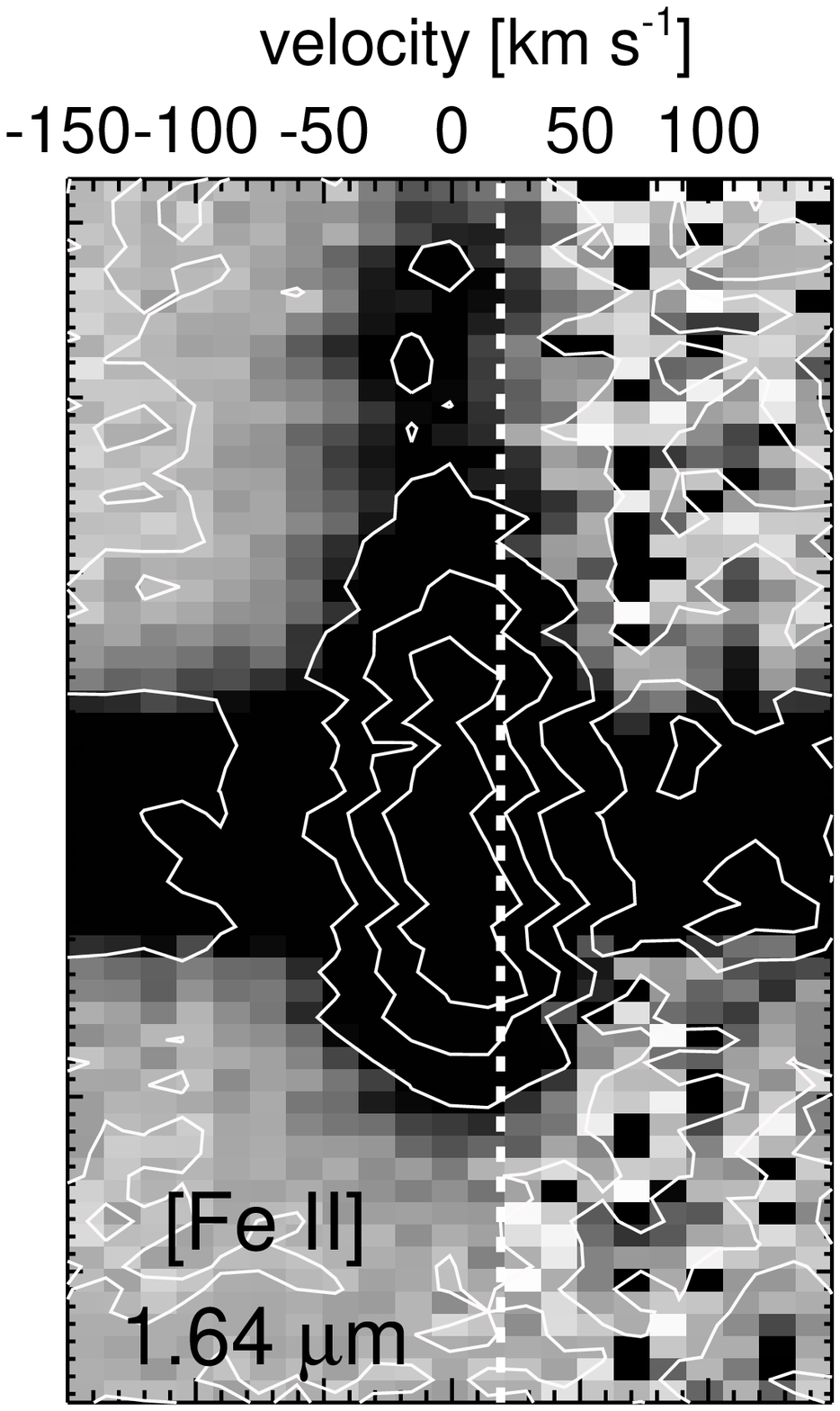} \\
\includegraphics[trim=20mm 0mm 10mm 10mm,angle=0,scale=0.325]{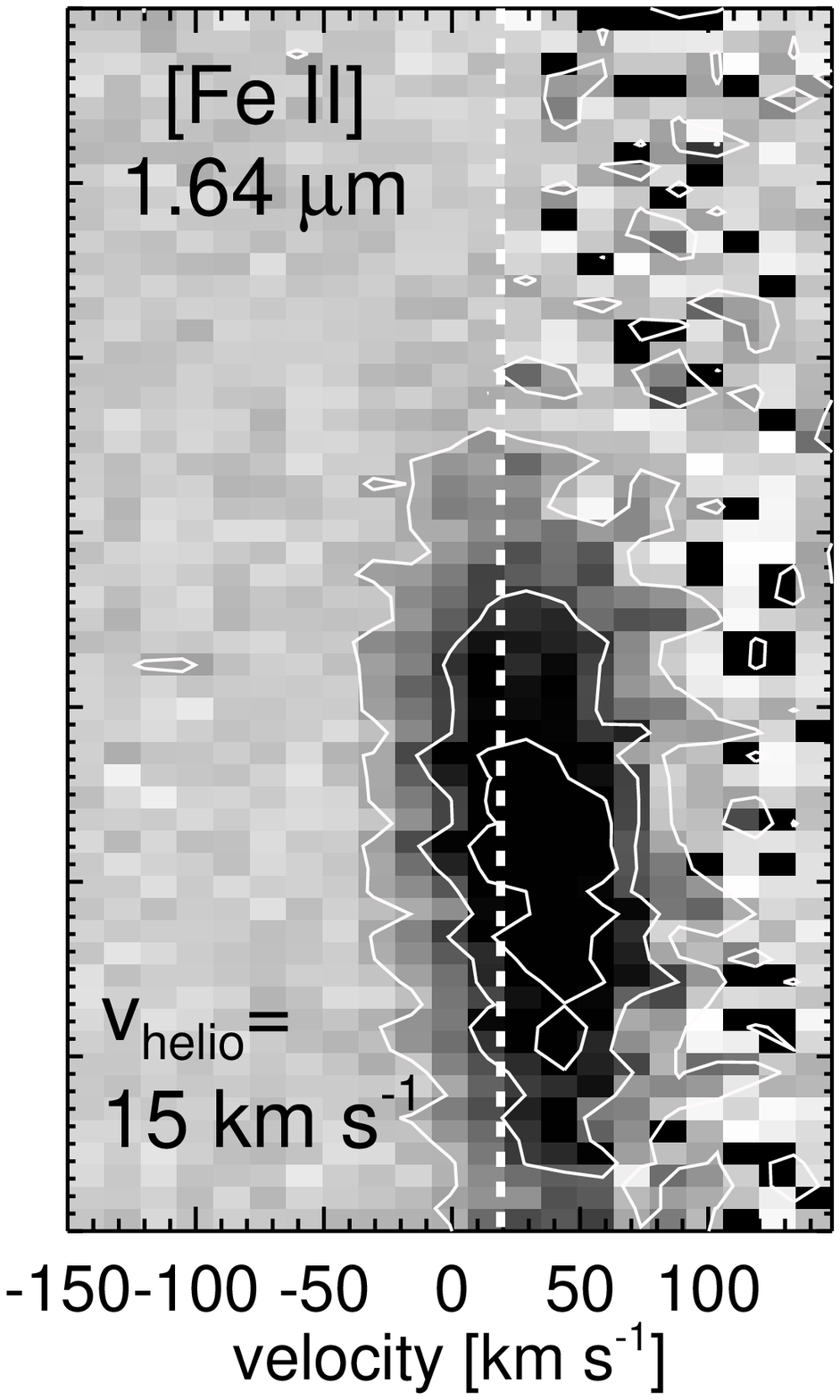} \\
\end{array}

\begin{array}{c}
\includegraphics[trim=20mm 10mm 15mm 0mm,angle=0,scale=0.325]{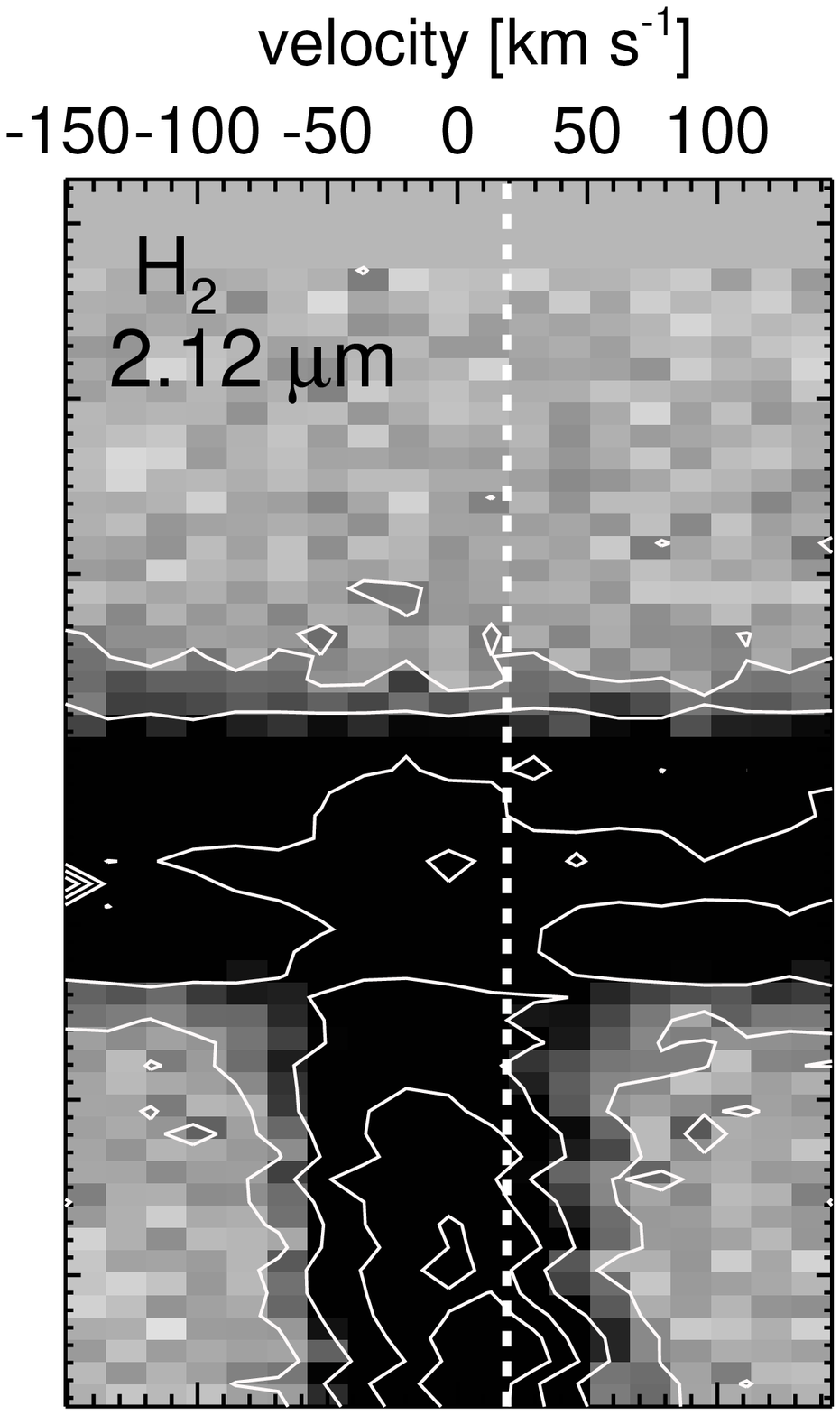} \\
\includegraphics[trim=20mm 0mm 15mm 10mm,angle=0,scale=0.325]{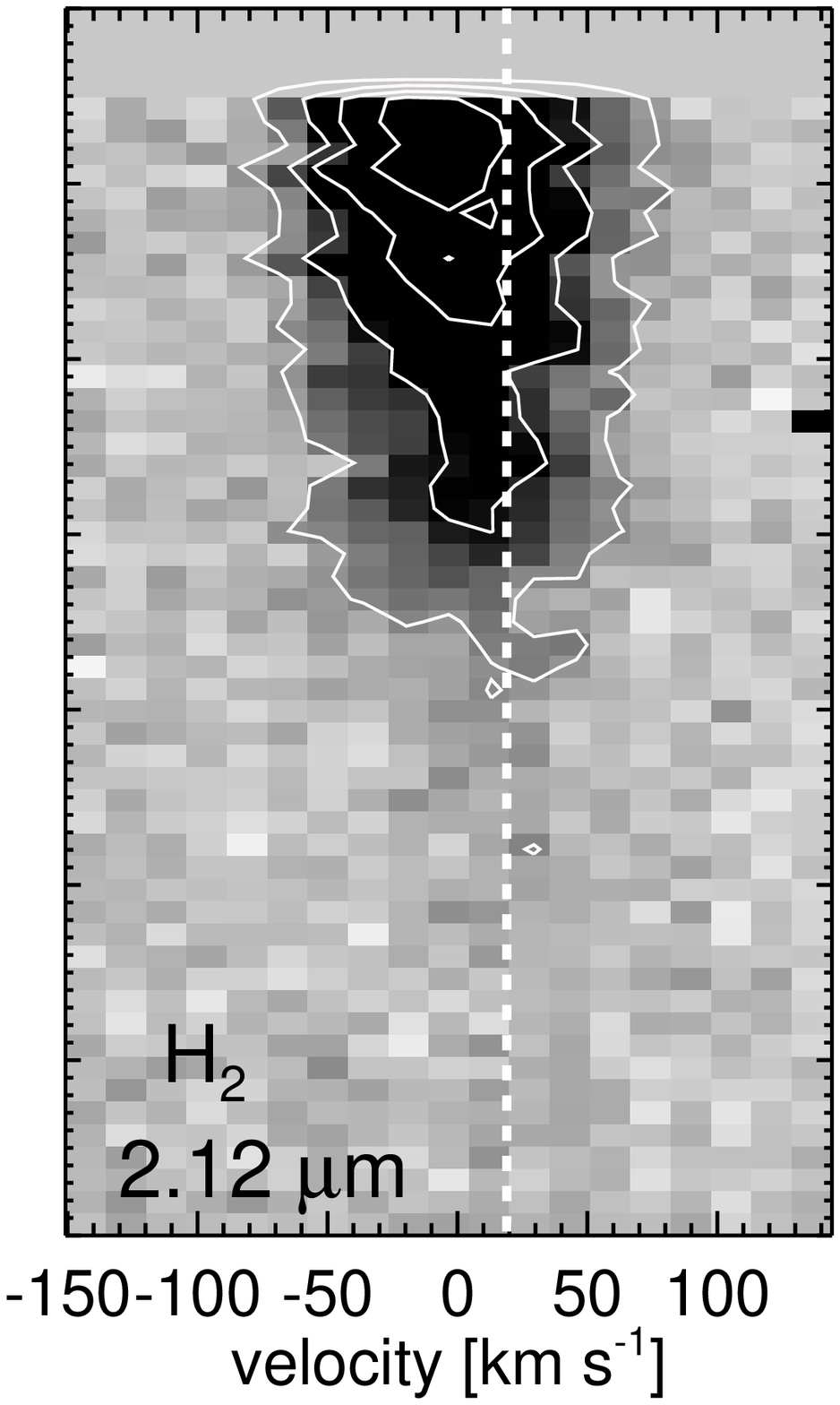} \\
\end{array}

\begin{array}{c}
\includegraphics[trim=15mm 10mm 0mm 0mm,angle=0,scale=0.325]{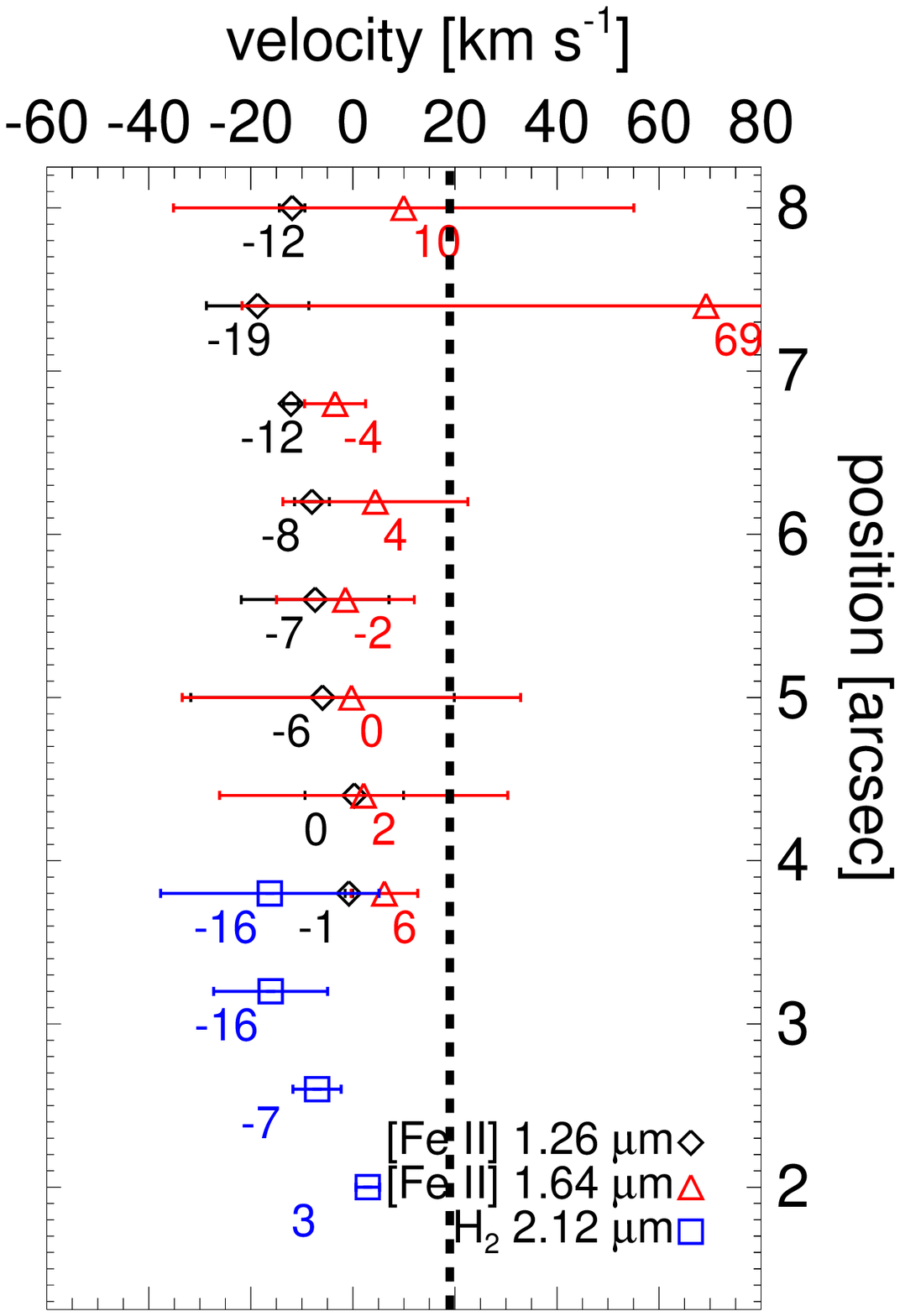} \\
\includegraphics[trim=15mm 0mm 0mm 10mm,angle=0,scale=0.325]{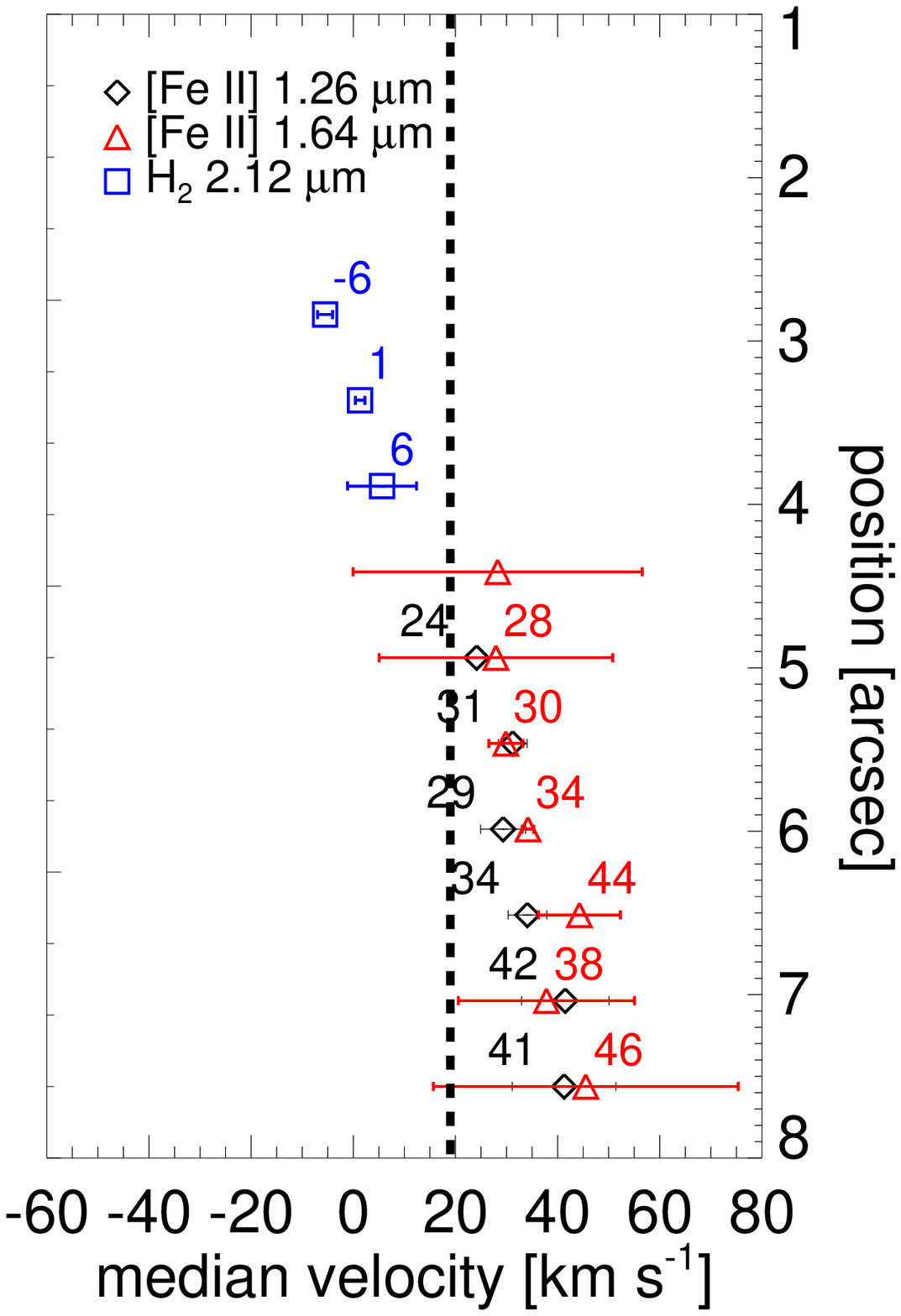} \\
\end{array}
\end{array}$
\caption{From left to right: 
FIRE slit positions drawn on a [Fe~{\sc ii}] 1.64 image, alongside 
position-velocity diagrams of HH~900 in the [Fe~{\sc ii}] 1.26 \micron, [Fe~{\sc ii}] 1.64 \micron, and H$_2$ S(0) 1-0 2.12 \micron\ lines, respectively in the heliocentric velocity frame. 
Plots on the far right show the median line velocity over steps of $\sim 0\farcs6$. 
Dashed lines indicate the systemic velocity inferred from velocity difference in the [Fe~{\sc ii}] emission from the eastern and western limbs of the jet. 
Continuum emission in spectra of the western limb of the jet (top row) is from the protostar in the middle of the putative H$\alpha$ microjet. 
[Fe~{\sc ii}] emission from the jet appears to be remarkably symmetrical on either side of the continuum. 
PV diagrams illustrate that the eastern and western limbs of the jet are redshifted and blueshifted (respectively) and the [Fe~{\sc ii}] emission grows bright where the H$_2$ emission fades away.  
}\label{fig:feii_pvs} 
\end{figure*}

\begin{figure*}
\centering
$\begin{array}{ccc}
\includegraphics[trim=15mm 0mm 0mm 0mm,angle=0,scale=0.495]{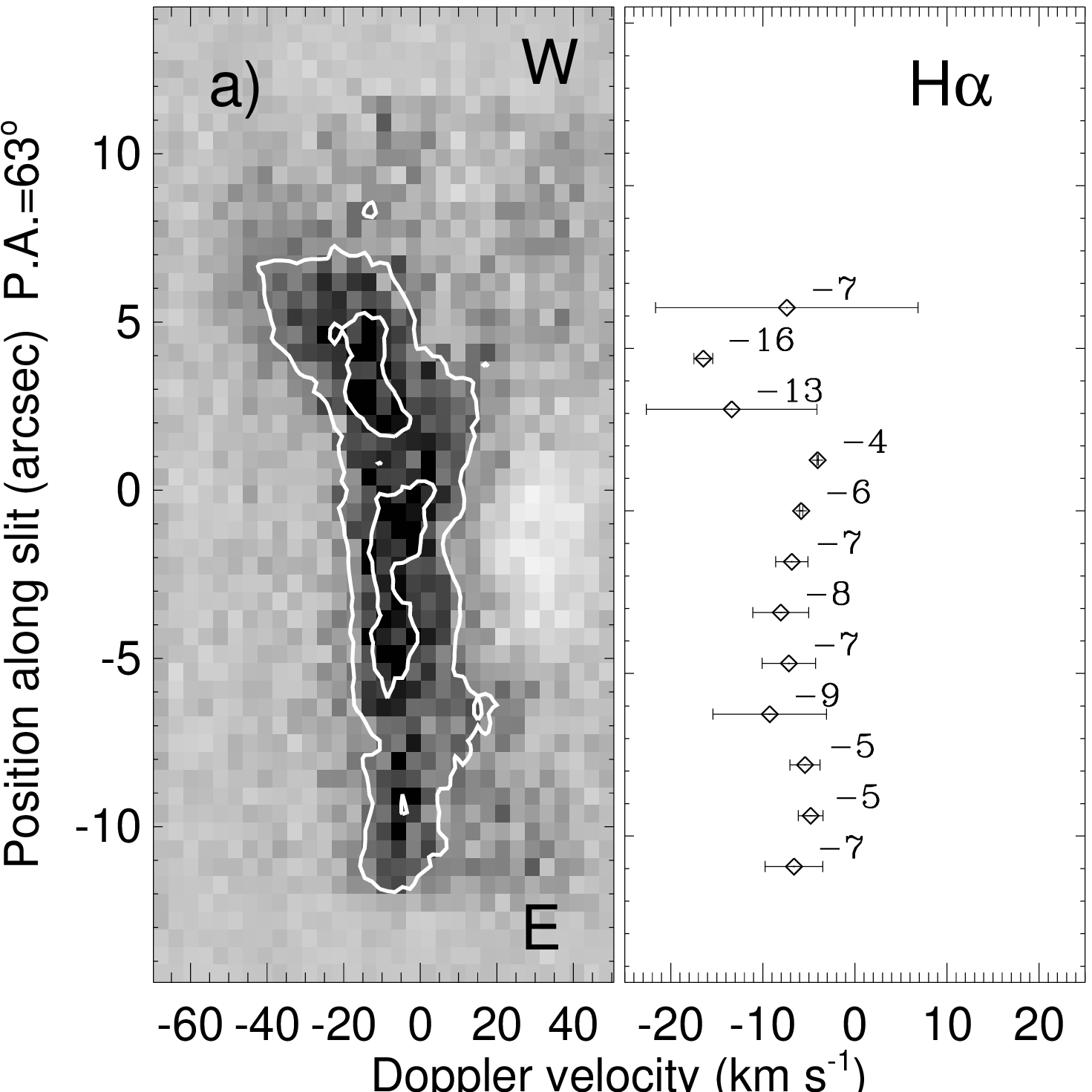} & 
\includegraphics[trim=0mm -26mm 15mm 0mm,angle=0,scale=0.275]{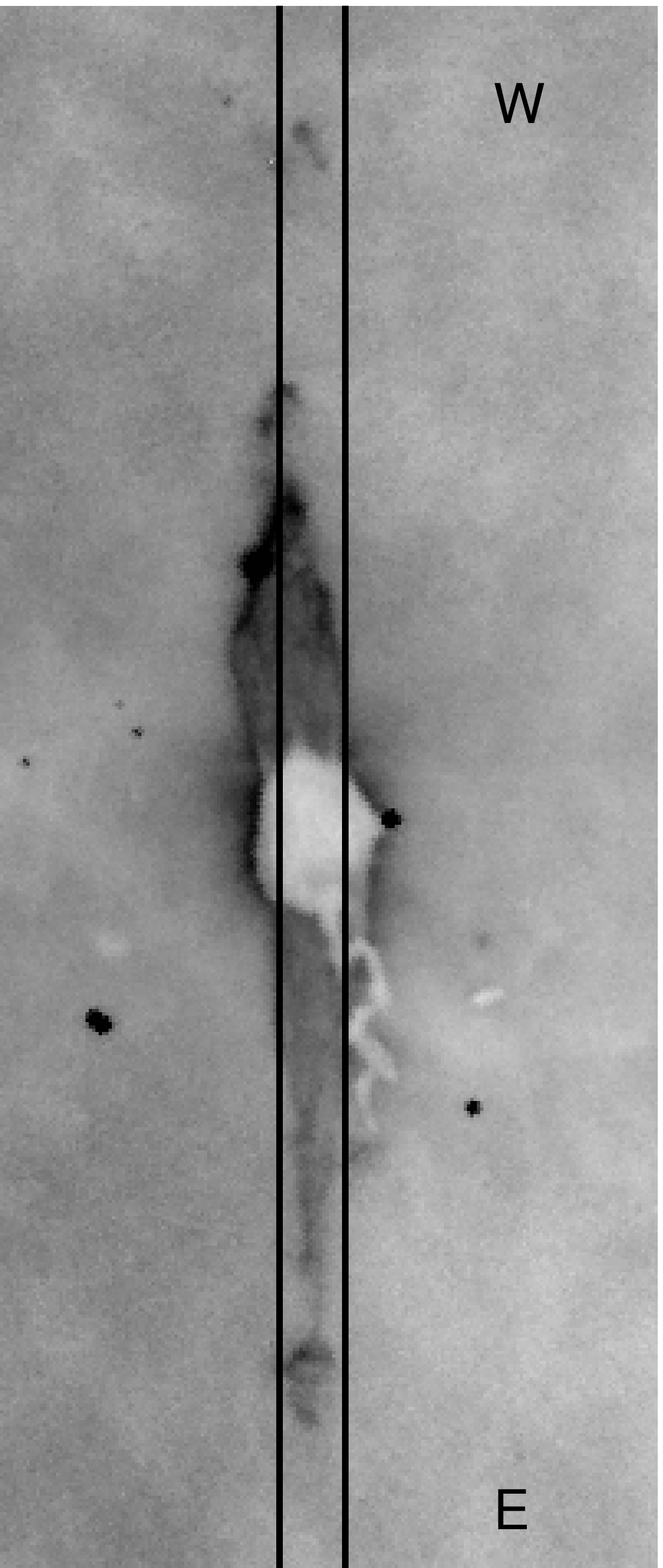} & 
\includegraphics[trim=10mm 0mm 0mm 0mm,angle=0,scale=0.495]{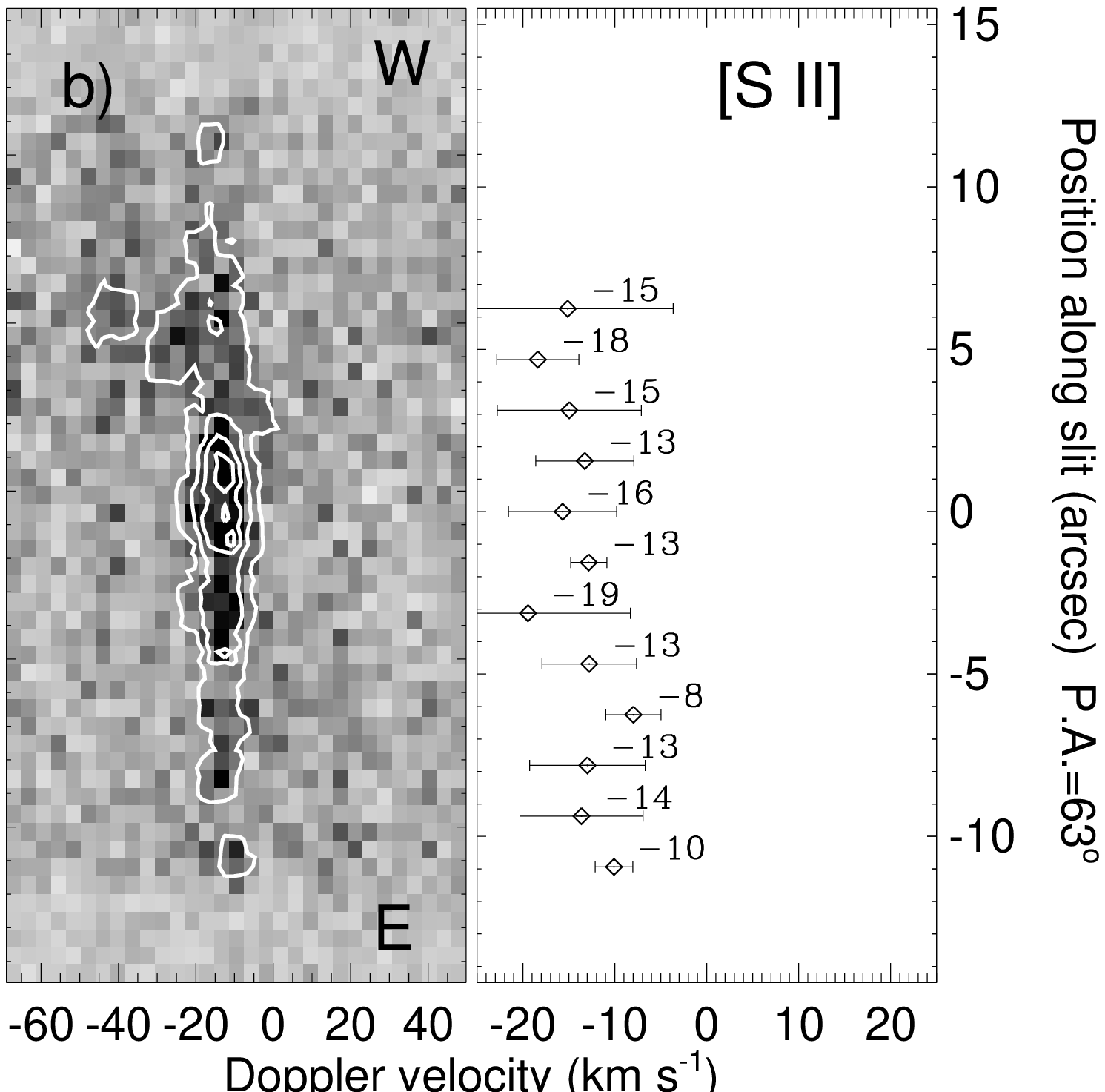} \\
\end{array}$
\caption{H$\alpha$ and [S~{\sc ii}] position-velocity diagrams of HH~900 shown on either side of an H$\alpha$ image with the EMMI slit position show at approximately the same spatial resolution. Plots show the median velocity taken over $\sim 2$\arcsec\ increments. 
}\label{fig:vis_pvs} 
\end{figure*}

Among the peculiarities of HH~900 revealed with H$\alpha$ images from \textit{HST} is the morphology of the dark globule. 
A small, wiggly, dark tail extends from the southeastern edge of the globule, creating a tadpole-like morphology (see Figure~\ref{fig:hst_ims}). 
The tail of the dark tadpole globule appears to lie in front of the jet, leading \citet{smi10} to suggest that the eastern limb of the jet is redshifted, and the western limb of the jet is blueshifted. 
Both optical and IR spectra now confirm this conjecture (see Figures~\ref{fig:feii_pvs} and \ref{fig:vis_pvs}). 
H$\alpha$ emission from the western limb of the jet is blueshifted and shows a Hubble-like velocity structure, with Doppler velocities extending up to roughly $-40$ km s$^{-1}$. 
A similar velocity structure is barely discernible in the [S~{\sc ii}] spectrum (the average of the $\lambda 6717$ and $\lambda 6731$ emission) due to the lower signal-to-noise. 

In contrast to the Hubble-like flow in the optical emission, near-IR [Fe~{\sc ii}] lines from both sides of the jet show a relatively constant velocity with redshifted velocities measured to be $\sim 30-60$ km s$^{-1}$ in the eastern limb of the jet and blueshifted velocities of $\sim 0 - 20$ km s$^{-1}$ in the western limb (suggesting $v_{sys} \approx 19$ km s$^{-1}$, see Figure~\ref{fig:feii_pvs}). 
The velocity difference between the eastern and western limbs of the jet is symmetric about the globule, indicating that the [Fe~{\sc ii}] emission is dominated by a single bipolar jet launched from within the globule. 
The H$_2$ emission from the eastern and western limbs of the jet is also redshifted and blueshifted, respectively. 
However, H$_2$ emission appears to be $\sim 20$ km s$^{-1}$ bluer than [Fe~{\sc ii}] velocities in both limbs. 
Position-velocity diagrams illustrate that the locations of H$_2$ and [Fe~{\sc ii}] emission from the jet appear to be mutually exclusive (see Figure~\ref{fig:feii_pvs}). 
In fact, H$\alpha$, [Fe~{\sc ii}], and H$_2$ emission all trace different velocity components (see Figures~\ref{fig:feii_pvs}~and~\ref{fig:vis_pvs}). 
Blueshifted velocities in the H$\alpha$ position-velocity diagram increase from $\sim0$ km s$^{-1}$ to the higher velocity traced by the [Fe~{\sc ii}] jet ($\sim -40$ km s$^{-1}$) while H$_2$ emission appears to be $\sim 20$ km s$^{-1}$ bluer than $v_{sys}$ inferred from the [Fe~{\sc ii}] emission from the innermost part of the jet.

\subsection{H$\alpha$ proper motions and 3-D velocities of jet features}\label{ss:pm}
With a $\sim 9$ yr time baseline between the two \textit{HST}/ACS epochs, we are sensitive to transverse velocities of $\sim 25$ km s$^{-1}$ in nebulous jet features seen in H$\alpha$, limited by the signal-to-noise of the jet knot compared to the bright background of the H~{\sc ii} region. 
We report proper motions and radial velocities in Table~\ref{t:pm}. 
Despite the relatively long time baseline, the morphology of the outflow has remained remarkably constant. 
Only the bow shocks and knots capping either side of the broad bipolar outflow show noticeable motion between the two epochs. 
Both knots move away from the globule along the jet axis with transverse velocities $\sim 100$ km s$^{-1}$ (see Figure~\ref{fig:pm}). 
The H$\alpha$-bright filament along the western limb of the jet does not appear to move at all ($<25$ km s$^{-1}$). 

The eastern and western bow shocks move in opposite directions, with proper motion velocities near the tip of the shock of $\sim 60$ km s$^{-1}$. 
For jets that lie near the plane of the sky (as appears to be the case for HH~900, see Table~\ref{t:pm}), the difference between the inclination-corrected proper motions of adjacent knots can be used to estimate the shock velocity. 
Both bow shocks have velocities $\sim 40$ km s$^{-1}$ slower than the inner jet, suggesting shock velocities $\sim 40$ km s$^{-1}$. 
Fast, dissociative J-shocks ($v_{shock} > 30-40$ km s$^{-1}$) are expected to be bright in near-IR [Fe~{\sc ii}] lines \citep{nis02,pod06} and indeed, both shocks have bright [Fe~{\sc ii}] emission (see Section~\ref{ss:morph} and Figure~\ref{fig:hst_ims}). 

Combining the transverse velocity of jet knots with the radial velocity from spectra, we can calculate the tilt angle, $\alpha = \mathrm{tan}^{-1} (v_r / v_T)$ and 3-D space velocity. With the low radial velocities in the H$\alpha$ spectrum, we find that the jet lies close to the plane of the sky (although we note the poor signal-to-noise of the spectrum). 
We derive a tilt angle of the jet away from the plane of the sky of $\lesssim 10^{\circ}$ (see Table~\ref{t:pm}).


\begin{table*}
\caption{HH~900 Proper Motions\label{t:pm}}
\centering
\begin{tabular}{rrrrrrrr}
\hline\hline
Object & $\delta$x & $\delta$y & v$_T$$^{\mathrm{a}}$ & v$_R$$^{\mathrm{b}}$ & 
velocity$^{\mathrm{c}}$ & $\alpha$ & age$^{\mathrm{d}}$ \\
 & mas & mas & [km s$^{-1}$] & [km s$^{-1}$] & [km s$^{-1}$] & 
[degrees] & yr \\ 
\hline
HH~900~A &   43 (1) &  -24 (0.4) &    60 (1) &  ... &   60 (1) & ... & 3523 (115) \\
HH~900~B &   31 (1) &  -2 (2) &    37 (2) &  ... &   37 (2) & ... & 5364 (307) \\
HH~900~C &   55 (0.1) &  -11 (1) &    68 (1) &  ... &   68 (2) & ... & 2803 (87) \\ 
HH~900~D &   74 (1) &  -31 (4) &    97 (3) &  ... &   97 (3) & ... & 859 (34) \\
HH~900~E &   76 (2) &  -25 (1) &    97 (3) &  ... &   97 (3) & ... & 682 (25) \\
HH~900~F &  -60 (1) &   42 (2) &    89 (2) &  -16 (4) &   90 (5) &  -10 (0.3) & 691 (24) \\
HH~900~G &  -77 (1) &   39 (1) &    104 (2) &  -7 (10) &   105 (10) &  -4 (0.1) & 746 (24) \\
HH~900~H &  -40 (3) &   14 (1) &    51 (3) &  ... &   51 (3) & ... & 5379 (365) \\
HH~900~I &  -42 (1) &   28 (1) &    61 (2) &  ... &   61 (2) & ... & 4664 (174) \\
PCYC~838 &  -3 (1) &   2 (1) &    4 (1) &  ... &   4 (1) & ... & 11201 (3126) \\
PCYC~842 &   5 (3) &  -11.8 (0.4) &    15 (2) &  ... &   15 (2) & ... & 1393 (140) \\ 
\hline
\end{tabular}
\begin{tabular}{l}
Proper motions measured for the YSOs and the knots marked in Figure~\ref{fig:pm}. \\
Uncertainties are listed in parentheses alongside the best-fit value. \\ 
$^a$ The transverse velocity, assuming a distance of 2.3 kpc. \\
$^b$ The radial velocity measured from spectra. \\
$^c$ The total velocity, assuming the average inclination 
when a radial velocity is not measured directly. \\
$^d$ Time for the object to reach its current location at the measured velocity, assuming ballistic motion. \\
\end{tabular}
\label{t:pm}
\end{table*}


\subsection{Protostar kinematics}\label{ss:yso_kin}
We can also constrain the motion of the two protostars near HH~900.
The YSO that lies along the western limb of the outflow (PCYC~838, see Figure~\ref{fig:hst_ims}) falls within the slit we use to observe the western limb of the jet. 
We extract the spectrum of this source separately. 
After subtracting off extended emission, we find that the central velocity of the hydrogen recombination lines in the spectrum of the YSO constrain the heliocentric Doppler velocity of the protostar to be $\lesssim |5|$ km s$^{-1}$ \citep[compared to the heliocentric systemic velocity of $-8.1\pm 1$ km s$^{-1}$ found for $\eta$ Car by][]{ns04}. 

\begin{figure}
\centering
\includegraphics[trim=25mm 50mm 25mm 50mm,angle=90,scale=0.325]{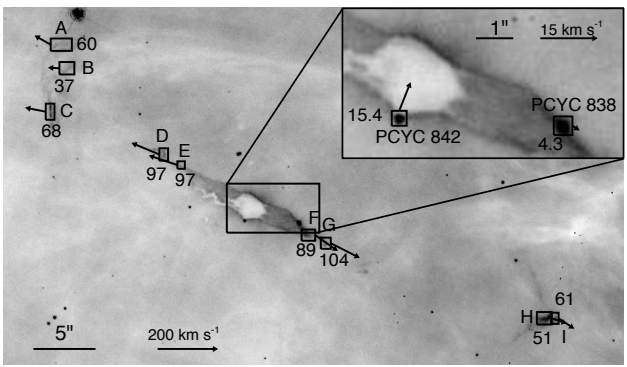} 
\caption{New \textit{HST}/ACS H$\alpha$ image of HH~900 with boxes used to measure proper motions overplotted. 
}\label{fig:pm} 
\end{figure}

We also measure the proper motions of both protostars. 
PCYC~838 moves in roughly the same direction as the jet with a transverse velocity of $4.3 \pm 1.3$ km s$^{-1}$ (see Table~2). 
Together, the radial and transverse velocities constrain the motion of PCYC~838 to $\lesssim 7$ km s$^{-1}$. 
We do not have spectra of the protostar near the bottom of the dark globule, PCYC~842 (see Figure~\ref{fig:hst_ims}), but we can measure its proper motion in the plane of the sky. 
The projected motion of PCYC~842 is \textit{toward} the globule with a transverse velocity of $15.4 \pm 3.0$ km s$^{-1}$ (see Figure~\ref{fig:pm}).

\section{Discussion}\label{s:discussion}

\subsection{Spatial offset of [Fe~{\sc ii}] emission}\label{ss:feii_offset}
Bright [Fe~{\sc ii}] emission on either side of the dusty tadpole globule in HH~900 traces a symmetric, narrow jet. 
Like HH~901 and HH~902, two externally irradiated jets that are also in the Carina nebula, the brightest [Fe~{\sc ii}] emission turns on more than 1\arcsec\ away from the edge of the globule, indicating a gap between the globule edge and the beginning of bright [Fe~{\sc ii}] emission from the jet. 
The offset of the [Fe~{\sc ii}] emission may be interpreted as evidence that the mass-loss rate in the jet has recently decreased, since a 
lower density jet may not maintain a column of material sufficient to shield Fe$^+$ from further ionization \citep{rei13}. 
However, HH~900, HH~901, and HH~902 all have gaps of similar size, and a nearly synchronized end of an outflow phase from three physically unrelated jets would seem highly suspicious. 
The greatest similarity between the three sources is their proximity ($\lesssim 3$ pc in projection) to a young massive cluster with log(Q$_{\mathrm{H}}$)$> 50$ photons s$^{-1}$ (Tr~14 for HH~901 and HH~902, Tr~16 for HH~900). 
This argues that the gap is produced by something in the environment, rather than due to a recent change in the mass-loss rate.

The three dark, narrow streaks in the western limb of HH~900 that \citet{smi10} interpret as dust filaments in the walls of the outflow cavity suggest another explanation for the offset [Fe~{\sc ii}] emission. 
These dusty streaks hint at a potentially large column of cold, dusty material that was blasted out of the globule by the jet. 
Close to the cloud edge, recently ejected dust and molecules may prevent UV emission from illuminating the inner jet.

We can measure reddening in the HH~900 jet using the ratio $\mathcal{R} = \lambda12567$/$\lambda16435$. 
$\mathcal{R}$ increases through the width of the jet (perpendicular to the direction of propagation), supporting the idea that external irradiation from massive stars in Tr~16 dominate the ionization and dust survival in the jet, not the nearby protostars or shocks. 
The increase in $\mathcal{R}$ also suggests that some entrained dust survives even in the outer regions of the jet.

\begin{figure}
\centering
$\begin{array}{|c|}
\hline
\includegraphics[trim=25mm 20mm 25mm 17.5mm,angle=90,scale=0.325]{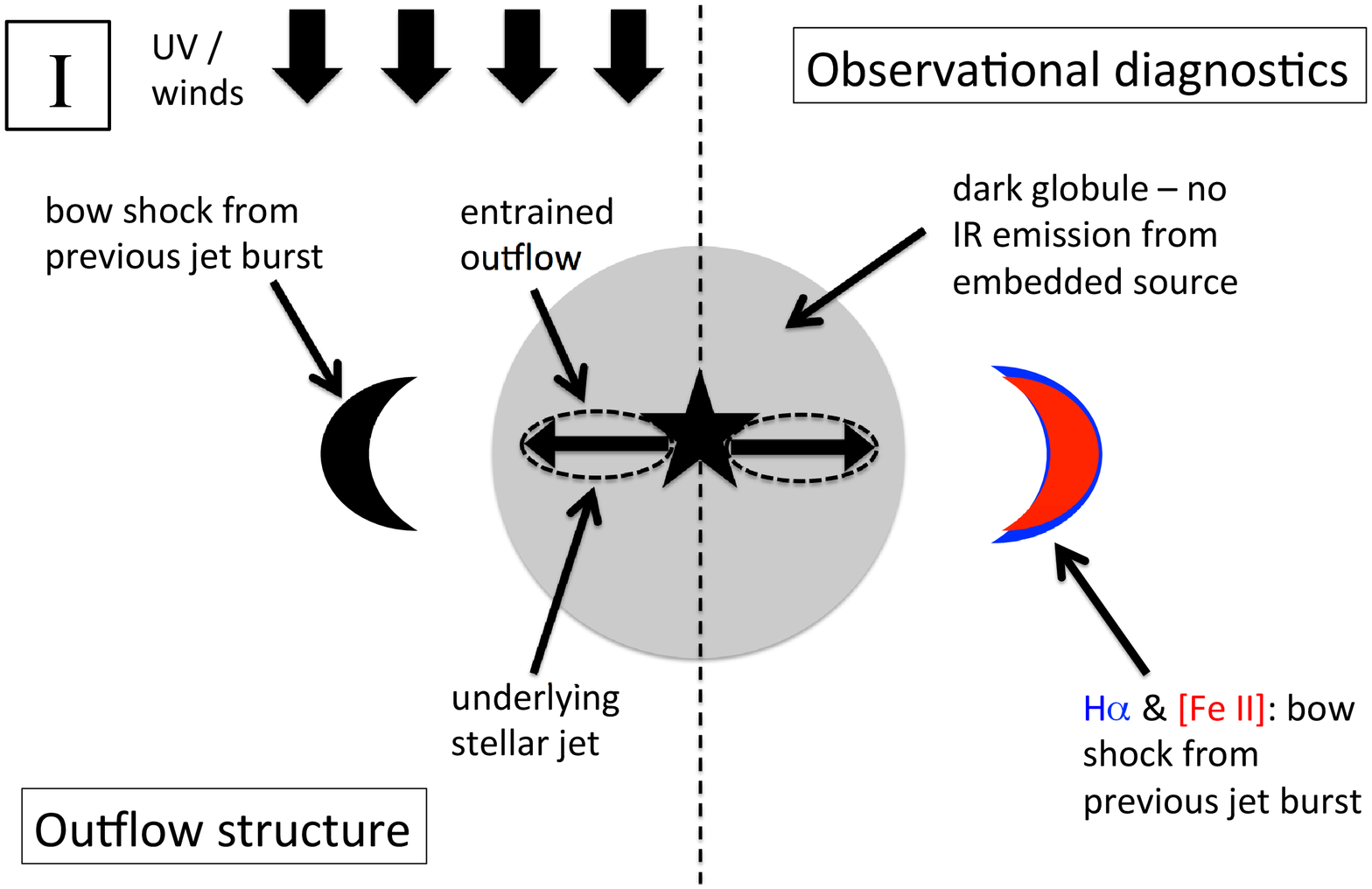} \\
\hline
\includegraphics[trim=25mm 20mm 25mm 17.5mm,angle=90,scale=0.325]{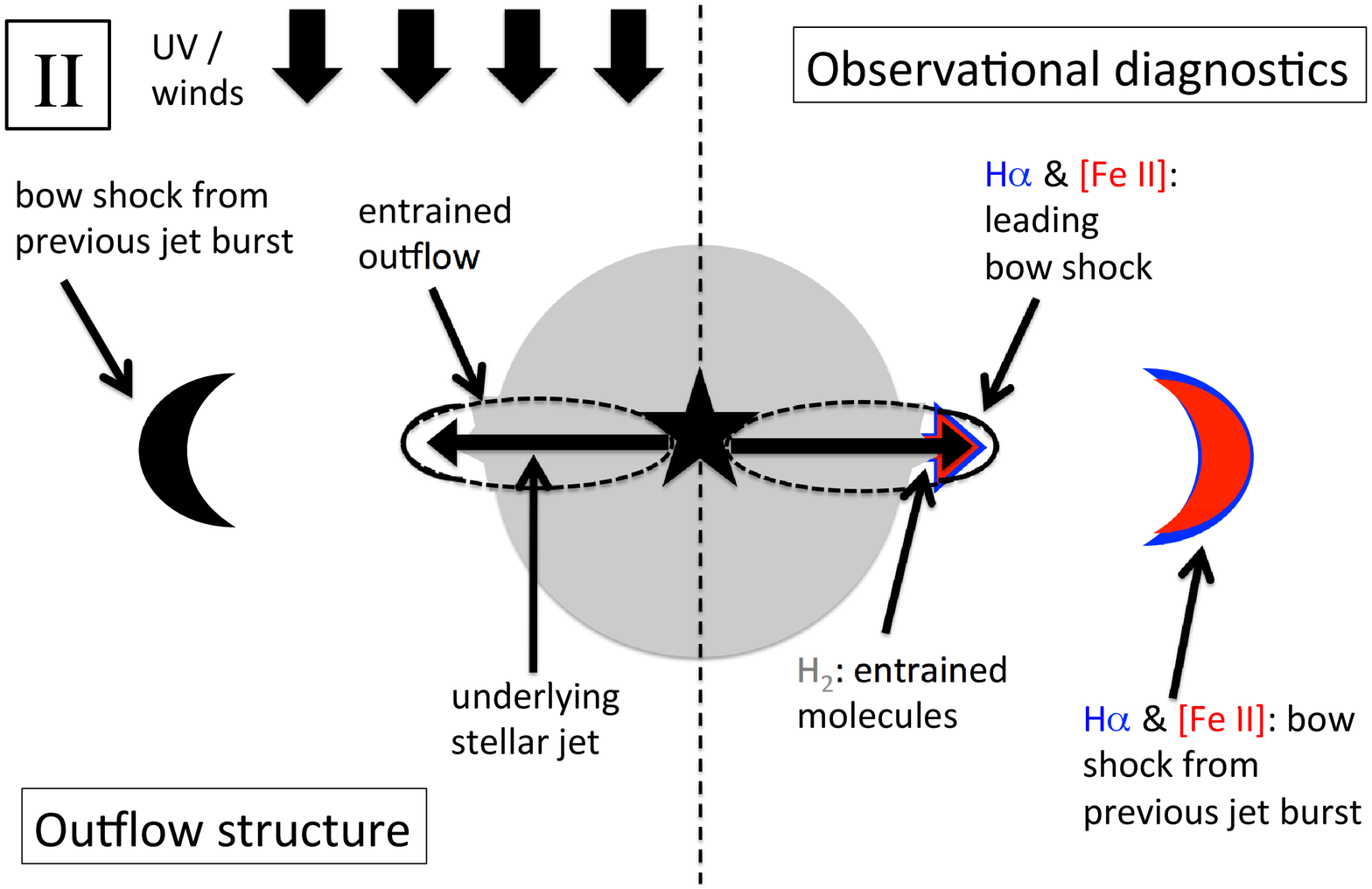} \\
\hline
\includegraphics[trim=25mm 20mm 25mm 17.5mm,angle=90,scale=0.325]{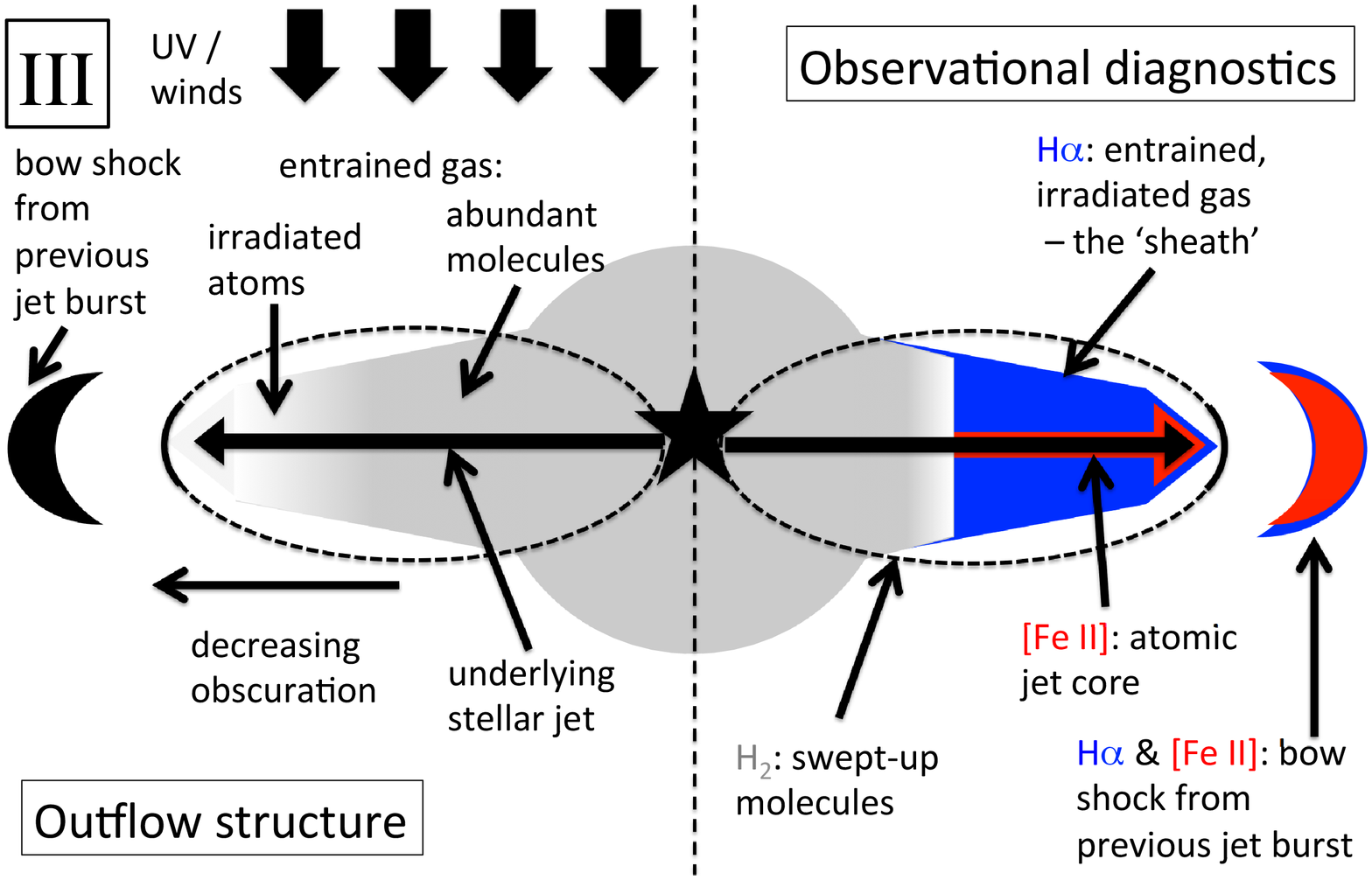} \\
\hline
\end{array}$
\caption{A time-series illustration of an outflow breaking out of a small, dense globule into the H~{\sc ii} region. 
Panel I shows the jet turning on and beginning to entrain an outflow before emerging from the globule as shown in panel II. Panel III illustrates the current geometry of the irradiated jet-outflow structure. 
Different materials will dominate the observed structure of the outflow as it propagates further from the globule and spends more time subject to external irradiation (left side). Observational tracers of the different components of the irradiated outflow (right side) reveal the decreasing obscuration of the jet and increasing impact of external irradiation. 
Red lines trace [Fe~{\sc ii}] emission, blue lines show H$\alpha$, and gray lines illustrate the location of molecules. 
}\label{fig:cartoon} 
\end{figure}

Dust entrained in the outflow may obscure the inner jet, explaining the offset between [Fe~{\sc ii}] emission and the edge of the globule. 
We propose that [Fe~{\sc ii}] emission in HH~900 traces the collimated jet while broad H$\alpha$ and H$_2$ emission near the globule trace the wider body of an externally illuminated outflow cavity (see Figure~\ref{fig:cartoon}). 
Narrow [Fe~{\sc ii}] emission tracing the body of the jet has been seen in other HH jets in Carina \citep[e.g. HH~666 and HH~1066, see][]{rei13}. 
Broad H$_2$ emission in HH~900 extends beyond the edge of the globule along the outflow direction, but only to the point where [Fe~{\sc ii}] emission begins, indicating that molecules persist in this intermediate zone. 
Extinction in the H$\alpha$ image also hints at the presence of entrained dust that survives in the H~{\sc ii} region close to the globule. 
Continuum-subtracted [Fe~{\sc ii}] and H$_2$ images are required to determine whether entrained dust and molecules also obscure the inner regions of HH~901 and HH~902.

HH~900 has two bow shocks separated from continuous emission in the inner jet, demonstrating that there has been a previous breakout of the jet into the H~{\sc ii} region. 
Despite this, new near-IR images show evidence for a significant column of dust and molecules outside the globule entrained by the jet. 
This may be expected from the youngest sources that are still surrounded by a substantial molecular envelope with only a small cavity cleared by the jet \citep{arc06}. 
Significant entrainment of molecular and dusty material by jets driven by more evolved protostars, or after many jet bursts, may be a direct consequence of the feedback-dominated environment (see Section~\ref{ss:enviro}). 
Compression of the molecular globule by massive star feedback may refill the cavity along the jet axis, increasing the amount of material entrained by a subsequent pulse of the jet.

Both HH~901 and HH~902 also show evidence for previous jet activity. 
If entrained material obscures the inner jet in all three cases, then this argues for a large column of material to be dragged out of the globule repeatedly by jet bursts. 
\citet{rei13} estimate $n_H \gtrsim 10^5$ cm$^{-3}$ for the pillar housing HH~901, and we expect similarly high densities for the HH~902 and HH~900 clouds (also see Section~\ref{ss:ysos_globs}). 
All three jets are embedded in the hot stellar wind bubble created by the many O-type stars in the nearby Tr~14/Tr~16 star cluster, so they are subject to similar amounts of feedback. 
Jets emerging from more diffuse globules \citep[e.g. HH~666, see][Reiter et al. in preparation]{smi04,rei13} do not show this same offset from the globule edge.

\subsection{The putative H$\alpha$ microjet}\label{ss:microjet}
Based on the morphology in the H$\alpha$ image, \citet{smi10} identify a possible microjet along the western limb of the broad HH~900 outflow. A bright point source with colours consistent with being a YSO lies in the middle of this bright H$\alpha$ filament and was identified as the possible driving source for the putative microjet. 
After examining the new imaging and spectroscopy presented here, we find it unlikely that this H$\alpha$-bright feature is a separate microjet. 
The key arguments against the microjet hypothesis are 
(1) the morphology in the [Fe~{\sc ii}] images, 
(2) radial velocities in spectra, 
and 
(3) proper motions as detailed below.

(1) The [Fe~{\sc ii}] emission indicates that the main body of the HH~900 jet is symmetric about the center of the globule and shows no significant deviation at the position of the putative microjet. 
[Fe~{\sc ii}] intensity tracings through both sides of the jet are similar, with no significant difference in the brightness of the western limb as compared to the east (see Figure~\ref{fig:feii_tracings} and Section~\ref{ss:morph}). 
It is possible that the microjet, driven by an unobscured star, is completely ionized, and therefore is only visible in H$\alpha$. 
However, if this were the case, the jet would have to be projected in front of the blueshifted limb of HH~900. 

(2) Linewidths of the [Fe~{\sc ii}] and H$_2$ emission from the western limb are similar to the eastern side of the jet, and appear to trace a single jet (see Figure~\ref{fig:feii_pvs} and Section~\ref{ss:velocity}). 
Moreover, velocity changes on either side of the jet are monotonic. 
There is no evidence for two jet velocity components in the H$\alpha$ spectrum (although the signal-to-noise and velocities are low, see Figure~\ref{fig:vis_pvs}). 

(3) The high degree of symmetry in the H$\alpha$ filament on either side of the protostar combined with the lack of any radial velocity changes suggests that if this feature is indeed a jet, it must lie almost exactly in the plane of the sky. 
However, our proper motion measurements constrain any oppositely directed, outward motions in the plane of the sky to $\lesssim 25$ km s$^{-1}$. 
Together, proper motions and spectra require that any outflowing gas must move slower than $\sim 30$ km s$^{-1}$, unusually slow for a jet driven by an unobscured protostar \citep[see, e.g.][]{rei01,rei14}.

Without evidence for any bipolar motion in the H$\alpha$ filament away from PCYC~838, it seems unlikely that this putative ``microjet'' is a separate jet at all. 
Hypothetically, the YSO that appears to lie at its center may have been ejected from the globule in a dynamical encounter, with the motion of the ejected star through the larger bipolar outflow creating a bright H$\alpha$ tidal tail that looks like an H$\alpha$ microjet. 
However, hydrogen emission lines in the spectrum of the YSO (seen intersecting the jet in Figure~\ref{fig:feii_pvs}) constrain the Doppler velocity of the protostar to be $v_{sys} \lesssim - 5$ km s$^{-1}$ whereas the H$\alpha$ emission at the same position is $\sim -12$ km s$^{-1}$. 
Moreover, we can constrain the proper motion of the YSO with the two epochs of \textit{HST}/ACS data. 
We find that the star moves almost imperceptibly in the H$\alpha$ images, with a measured velocity in the plane of the sky of $4.3 \pm 1.3$ km s$^{-1}$. 
The YSO is not moving faster than the random motion of stars in the Carina nebula, making the ejection scenario unlikely. 
At these low velocities, the dynamical time required for the star to reach its current position is $\sim 11,000$ yr, longer than the estimated dynamical age of the two outer bow shocks in HH~900 ($\sim 2200$ yr, see Section~\ref{ss:morph}), and much longer than the age of the inner jet ($\sim 1000$ yr, see Section~\ref{ss:ysos_globs}). 
Given the discrepancy in the timescales for the star and the jet, any interaction between the star and HH~900 is likely due to the jet recently moving past the position of the YSO. 
The simplest explanation is a chance alignment of the YSO with the apparent center of the H$\alpha$-bright filament. 
Indeed, \citet{wol11} find a relatively high spatially averaged source density in the Tr~16 subcluster near HH~900 (45 src pc$^{-2}$ in sub-cluster 12).

The H$\alpha$-bright filament may result from local irradiation due to the small separation between the YSO and the HH~900 outflow. 
Local irradiation from the driving protostar has been invoked to explain bright optical and IR emission from the inner regions of protostellar jets in more quiescent regions \citep[e.g.][]{rei00}. 
The IR-bright YSO that appears to lie in the middle of this filament must lie extremely close to the broad, bipolar outflow ($\lesssim 4.5 \times 10^{-4}$ pc or $\lesssim92$ au in the plane of the sky). 
Uncertainties in the three-dimensional geometry of the system may hide a slightly greater separation, although the presence of especially bright H$\alpha$ emission on either side of the YSO suggests that the protostar may affect the larger, apparently separate outflow. 
Thus, while our kinematic data rule out the possibility that this bright H$\alpha$ feature is a separate microjet, its origin remains elusive.

\subsection{Mass-loss rate in the neutral jet}\label{ss:mdot}

\citet{smi10} estimated mass-loss rates, $\dot{M}_{jet}$, of the jets in Carina using the H$\alpha$ emission measure in ACS images. 
For HH~900, the wide inner jet in the east and the H$\alpha$-bright filament (the now refuted microjet) in the west represent the two highest $\dot{M}_{jet}$ in the sample. 
However, mass-loss rates calculated from the H$\alpha$ emission measure reflect the total $\dot{M}_{jet}$ only for fully ionized jets, and otherwise provide a lower-limit. 
\citet{rei13} argue that the HH jets in Carina are not fully ionized, and that [Fe~{\sc ii}] emission traces neutral material in jets that are sufficiently dense to self-shield (i.e. to prevent Fe$^+ \rightarrow$ Fe$^{++}$). 
Relaxing the assumption that the jets are fully ionized and calculating the density required to self-shield generally increases the estimated mass-loss rate by at least an order of magnitude \citep{rei13}. 
For HH~900, the situation is unfortunately even more complex. 
H$\alpha$ and [Fe~{\sc ii}] have different spatial distributions and appear to trace different velocity components, suggesting that the mass-loss rate estimated from H$\alpha$ only samples the mass contained in a thin ionized skin on the outside of the outflow sheath entrained by the [Fe~{\sc ii}] jet. 
This is remarkable since the mass-loss rate derived from the H$\alpha$ emission measure was already the highest of all the jets in Carina \citep{smi10}. 

We can estimate the mass-loss rate of the HH~900 [Fe~{\sc ii}] jet by requiring that a neutral, cylindrical jet survives out to a distance $L_1$ before it is completely photoablated \citep{bal06,rei13}. 
Here, we take $L_1$ to be the length of the continuous [Fe~{\sc ii}] jet (not including the bow shocks). 
For a cylindrical jet, 
\begin{equation}
\dot{M}_{jet} \approx \frac{L_1 f \mu m_H c_{II}}{2D} \left[ \frac{\alpha_B}{\pi r_{jet} Q_H sin(\beta)} \right]^{-1/2} 
\end{equation}
where 
$f \approx 1$ is the filling factor, 
$\mu \approx 1.35$ is the mean molecular weight, 
$m_H$ is the mass of hydrogen, 
$c_{II} \approx 11$ km s$^{-1}$ is the sound speed in ionized gas, 
and 
$\alpha_B \approx 2.6 \times 10^{-13}$ cm$^3$ s$^{-1}$ is the Case B recombination coefficient. 
CPD-59~2641 is an O5~V star located $\sim1$\arcmin\ from HH~900, or a projected  distance $D\sim 0.7$ pc (which may underestimate the true separation by $\sim \sqrt{2}$). 
Assuming that CPD-59~2641 dominates the ionization, the ionizing photon luminosity is log$(Q_H)=49.22$ s$^{-1}$ \citep{smi06}. 
We measure the jet radius, $r_{jet} \approx 0.007$ pc, from the [Fe~{\sc ii}] images and assume that the angle $\beta$ between the jet axis and the direction of the ionizing radiation, is $\beta \approx 90^{\circ}$. 
From this, we find $\dot{M}_{jet} \gtrsim 7 \times 10^{-6}$ M$_{\odot}$ yr$^{-1}$ and $\dot{M}_{jet} \gtrsim 5 \times 10^{-6}$ M$_{\odot}$ yr$^{-1}$, respectively, for the eastern and western limbs of the jet \citep[compared to $5.68 \times 10^{-7}$ M$_{\odot}$ yr$^{-1}$ and $6.20 \times 10^{-7}$ M$_{\odot}$ yr$^{-1}$, respectively, estimated from the H$\alpha$ emission measure, see][]{smi10}.

As \citet{rei13} found for other HH jets in Carina, using [Fe~{\sc ii}] as a diagnostic and without assuming the jet is fully ionized, $\dot{M}_{jet}$ is an order of magnitude larger than \citet{smi10} estimate from the H$\alpha$ emission measure. 
Assuming that the mass \textit{accretion} rate, $\dot{M}_{acc}$, onto the protostar is $\sim 10-100$ times larger than the mass-loss rate in the jet \citep[e.g.][]{har95,cal98}, we find $\dot{M}_{acc} \approx 10^{-4}$ M$_{\odot}$ yr$^{-1}$. 
This points to a high luminosity from the driving protostar, either due to its relatively high mass or a recent accretion burst. 

\subsection{Protostars and globule survival}\label{ss:ysos_globs}

The jet axis defined by the collimated [Fe~{\sc ii}] emission rules out the hypothesis that the driving source of the jet is the protostar at the bottom of the HH~900 globule \citep[identified by][]{shi13}. 
Although the jet axis defined by the [Fe~{\sc ii}] emission in the WFC3-IR images from \textit{HST} indicates a clear mis-match of the star and the jet (see Figure~\ref{fig:fe_contours}), is it possible that the star began in the globule and was ejected, perhaps in a small-N stellar interaction \citep[e.g.][]{rei10,rei12}? 
If this is the case, the star must have been in the globule at the time that the innermost jet emission we detect in [Fe~{\sc ii}] was ejected. 
Assuming a jet velocity of 100 km s$^{-1}$ for the [Fe~{\sc ii}] emission located $\sim 1\farcs5$ from the globule edge (thus $\sim 2\farcs5$ from the presumed position of the driving source), this means the star would need to have been ejected within the last $\sim 275$ yr. 
To travel to its current location at the bottom of the globule in that time, the star would have to be moving at $\gtrsim 60$ km s$^{-1}$. 
However, PCYC~842 has an observed proper motion of $15.4$ km s$^{-1}$ \textit{toward} the globule (see Section~\ref{ss:yso_kin} and Table~\ref{t:pm}), clearly inconsistent with the ejection hypothesis. 

Without compelling evidence for the recent ejection of the jet-driving source, we must conclude that an additional protostar remains hidden inside the dark globule. 
Symmetric [Fe~{\sc ii}] emission from the eastern and western sides of the jet, including a similar offset from the globule edge (see Figure~\ref{fig:feii_tracings}), argues that the driving source is located inside the globule. 
Dark lanes in the blueshifted western limb of the jet apparent in the H$\alpha$ image connect the jet directly with the globule and require that it is driven by a protostar embedded within it. 
However, we do not detect IR emission associated with a protostar in the globule.

The high mass-loss rate in HH~900 makes the non-detection of a protostar particularly puzzling. 
\citet{smi10} and \citet{rei13} have argued that the HH jets in Carina are driven by intermediate-mass ($\sim 2-8$ M$_{\odot}$) protostars based on their high mass-loss rates, and the luminosity of candidate driving sources found for some of the jets support this interpretation \citep{smi04,ohl12,rei13,rei14}. 
Even if the HH~900 driving source were a low-mass protostar, it would need to be undergoing an FU-Orionis-type outburst to create a jet of this strength \citep{cro87,cal93}, and would therefore likely have a high accretion luminosity. 
However, we note that the dynamical age of the inner jet is $\sim 1000$ years (for a jet velocity of $100$ km s$^{-1}$), a factor of $\sim 10$ longer than the typical decay time of an FU~Orionis outburst \citep{hk96}. 
Small velocity differences between the redshifted and blueshifted sides of the jet suggest that the outflow axis lies close to the plane of the sky, and we find a small inclination angle for the jet($\alpha \lesssim 10^{\circ}$). 
Thus the circumstellar disc of the driving source will be seen almost edge-on. 
A YSO viewed through the midplane of an optically thick disc will be heavily extincted by a large column of circumstellar material. 
If this YSO is embedded in a dense and sufficiently opaque globule, no scattered light will escape the cloud and the protostar will remain unseen. 

To estimate the amount of circumstellar material required to completely obscure the driving source, we use the YSO models from \citet{rob06}. 
Since we have only upper limits on the protostar postulated to be in the globule (see Section ~\ref{ss:ysos}), we instead do a parameter search of the \citet{rob06} models.   
We make the conservative estimate that the protostellar mass is $1.95 - 2.05$ M$_{\odot}$ and the opening angle of the envelope cavity angle is $5-10^{\circ}$. 
We estimate a small cavity opening angle based on the fact that H$\alpha$ and H$_2$ emission are not wider than the globule. 
Within this search criteria, we sample disc accretion rates as high as $\sim 10^{-5}$ M$_{\odot}$ yr$^{-1}$, on the lower end of the accretion rates we expect if $\dot{M}_{jet} \approx 0.01 - 0.1 \times \dot{M}_{acc}$ (see Section~\ref{ss:mdot}). 
The disc accretion rate will be even higher (leading to a higher accretion luminosity) if the driving source is more massive (closer to 8 M$_{\odot}$), or undergoing an FU Orionis outburst \citep[with disc accretion rates $\gtrsim 10^{-4}$ M$_{\odot}$ yr$^{-1}$, e.g.][]{hc95,cal98}. 
For the 12 models that satisfy this search criteria, the estimated line-of-sight extinction to the stellar surface ranges from $A_v \approx 5 \times 10^3 - 2 \times 10^7$. 
Allowing for larger cavity opening angles of $40-45^{\circ}$ corresponding to more evolved driving sources, $A_v$ may be as low as $\approx 3 \times 10^2$. 
Taking the lower bound, $A_v = 10^2$, we can make a crude estimate of the globule mass. 
Using $N_H \approx 1.8 \times 10^{21} \times A_v$ cm$^{-2}$ \citep{guv09}, the column density along the line of sight must be at least $\sim 2 \times 10^{23}$ cm$^{-2}$. 
Assuming the globule is a uniform sphere, we can estimate the volume density. 
The diameter of the globule is $\sim 2$\arcsec, corresponding to a linear size of $\approx 0.02$ pc at the distance of Carina. 
For a globule radius of $0.01$ pc or $3\times10^{16}$ cm, this rough estimate yields an average volume density $n \sim 6 \times 10^6$ cm$^{-3}$, among the higher volume densities inferred from dust emission of low-mass star forming cores ($\sim 10^4 - 10^6$ cm$^{-3}$, \citealt{eno07,eno08}, although higher density cores have also been observed, e.g. \citealt{tok14}). 
Taking the globule to be a uniform sphere, this density suggests a globule mass of $\gtrsim 1$ M$_{\odot}$.

\citet{gre14} use the H$\alpha$ images of \citet{smi10} to estimate the extinction in the HH~900 globule and calculate a mass of $\sim 14$ M$_{\mathrm{Jupiter}}$, two orders of magnitude smaller than our estimate.  
This single-wavelength assessment of the extinction provides only a lower limit on the true column of material in the globule and is compromised by emission from the surface of the limb-brightened globule. 
Indeed, requiring that the globule has survived photoevaporation from the many O-type stars in Tr 16 for $\gtrsim 3.5$ Myr \citep{smi06} argues for a much larger mass reservoir. 
The photoevaporation mass-loss rate of the globule is 
\begin{equation}\label{eq:mdotphot}
\dot{M}_{phot} \approx 4 \pi r^2 n_H \mu m_H c_{II}
\end{equation}
where $r$ is the radius of the globule, 
and $n_H$ is the density of neutral hydrogen. 
We can estimate the density at the ionization front from the requirement that ionizations are balanced by recombinations. 
This gives 
\begin{equation}
n_{IF} \approx \sqrt{ \frac{Q_H}{8 \pi D^2 r \alpha_B} } 
\end{equation}
where, as in Section~\ref{ss:mdot}, we assume that the nearby O-type star CPD-59~2641 dominates the ionization. 
This yields $n_{IF} \sim 4000$ cm$^{-3}$. 
We can connect this to the density in the globule using the continuity equation, $n_{IF}c_{II} = n_H c_I$ where $c_I \approx 3$ km s$^{-1}$ is the sound speed in the molecular globule. 
For the estimated $n_H \sim 1.5 \times 10^4$ cm$^{-3}$, we find $\dot{M}_{phot} \sim 7 \times 10^{-6}$ M$_{\odot}$ yr$^{-1}$. 
This is within a factor of two of the photoevaporation rate \citet{smi04b} calculate for the nearby finger globule in Carina ($\sim 2 \times 10^{-5}$ M$_{\odot}$ yr$^{-1}$) as the higher flux of the ionizing source (log$(Q_H)=50.0$ s$^{-1}$) is roughly canceled by its larger separation ($\sim 6$\arcmin). 
With this high photoevaporation rate, a 14 M$_{Jup}$ globule will rapidly be ablated, and completely destroyed in $\sim 2000$ years. 
In contrast, the remaining lifetime of a $\gtrsim 1$ M$_{\odot}$ globule will be $\gtrsim 10^5$ years.

\subsection{Comparison to molecular outflows}\label{ss:mol_outflows}
\citet{rei13} postulated that the irradiated jets in Carina are similar to the jets that drive molecular outflows, but are laid bare by the harsh UV radiation permeating Carina. 
In this picture, molecules entrained in the outflow that are swept into the H~{\sc ii} region will be quickly dissociated, and should only be observed within a few arcsec of the edge of the natal globule.  
Indeed, H$_2$ emission in HH~900 survives only $\sim 1.5$\arcsec\ away from the edge of the molecular globule, and must have been introduced into the H~{\sc ii} region recently. 
Other authors have invoked a two-component outflow model to explain the coexistence of a high velocity jet and a slower, broader molecular outflow created by a wide-angle wind (e.g. HH~111, \citealt{nag97,lee00}; HH~315, \citealt{arc02}; HH~46/47, \citealt{arc13}).

In HH~900, there is the further complication that the H$\alpha$ position-velocity diagram traces a smooth increase in velocity with increasing distance from the dark globule. 
In the case of HH~111, higher velocity emission far from the driving source has been seen in maps of the CO emission and interpreted as evidence of jet-bow-shock entrainment \citep[e.g.][]{lee00,lef07}. 
As a bow shock propagates through its parent cloud, it will sweep up material, creating a sheath that envelopes the jet traveling in its wake. 
The initial impulse of the jet turning on (or up) creates the Hubble-like velocity structure. 
A corresponding shell of material swept up in HH~900 will be externally irradiated by Tr~16 upon exiting the globule, creating an ionized outer layer that will emit H$\alpha$. 
Within $\sim 1.5$\arcsec\ of the globule, this ionized sheath still exhibits H$_2$ emission. 
The overall blueshift of the H$_2$ velocities ($\sim 20$ km s$^{-1}$) compared to [Fe~{\sc ii}] suggests that H$_2$ emission is dominated by the expansion of the outflow cavity. 
The abrupt end of H$_2$ emission where [Fe~{\sc ii}] emission begins may signal a transition in the column of obscuring, entrained material from a mix of atoms and molecules to primarily atomic gas. 
Beyond this point, H$\alpha$ emission narrows until it converges with the [Fe~{\sc ii}] morphology at the head of the inner jet, as expected for prompt entrainment. 
Indeed, the Hubble-like velocities in the H$\alpha$ position-velocity diagram suggest that HH~900 is the irradiated analog of an entrained molecular outflow.

\subsection{Environmental interaction}\label{ss:enviro}

Theoretical work on small globules in H~{\sc ii} regions has explored their photoevaporation and the possibility that radiative compression may trigger star formation \citep[e.g.][]{ber89,ber90,ber96,gor02,dal07a,dal07b}. 
The small size of HH~900 and its survival despite its proximity to $\eta$ Car hint that the protostar forming within it may have been triggered by radiation-driven implosion. 
There remains an ambiguity between the triggering of new star formation and the uncovering, and/or possible acceleration, of star formation that would have happened anyway \citep{dal07b}. 
However, in feedback dominated environments, back-pressure from the photoevaporative flow off the globule will continue to compress the globule even as the protostar evolves. 
From detailed study of the `defiant finger,' another small globule in the Carina nebula, \citet{smi04b} find that back-pressure from the photoevaporative flow is a factor of $\sim 5$ greater than thermal support in the globule. 
If similar conditions apply to HH~900, then pressure from the ionization front may alter the physical conditions of the local star forming environment. 
By increasing the local pressure, feedback from massive stars may affect the envelope accretion rate, the amount of molecular and dusty material entrained by the jet (as well as the degree that it can clear an outflow cavity in the molecular envelope), and the rate at which the developing protostar destroys the globule from within. 
Clearly, hydrodynamic models of the evolution of protostars in irradiated globules would be interesting.

\section{Conclusions}\label{s:conclusion}

We present new imaging and spectroscopy of HH~900, a peculiar protostellar outflow in the Carina nebula. 
H$\alpha$ images reveal an unusually wide-body bipolar outflow while new narrowband [Fe~{\sc ii}] 1.26 \micron\ and 1.64 \micron\ images of HH~900 obtained with WFC3-IR trace a collimated jet. 
[Fe~{\sc ii}] emission remains remarkably symmetric on either side of the globule, suggesting that the unseen driving source lies roughly at the center of the globule. 
New narrowband H$_2$ images from GSAOI reveal extended molecular emission from the outflow just outside the globule that disappears at the same separation where the [Fe~{\sc ii}] emission from the jet grows bright ($\sim 1.5$\arcsec). 
The H$_2$ $\rightarrow$ [Fe~{\sc ii}] transition zone is seen on both sides of the outflow and is roughly symmetric. 
Both H$\alpha$ and H$_2$ emission are as wide as the globule when they emerge into the H~{\sc ii} region. 
Their width and kinematics suggest that these lines trace dusty and molecular material entrained by the jet that was recently dragged into the H~{\sc ii} region. 

Optical and near-IR spectra reveal separate and distinct velocity components traced by [Fe~{\sc ii}], H$_2$, and H$\alpha$ emission. 
Both [Fe~{\sc ii}] lines trace the steady jet emission from the red- and blueshifted (eastern and western, respectively) limbs of the jet. 
On the other hand, velocities in the H$\alpha$ and H$_2$ spectra increase further away from the globule, similar to the Hubble-like velocity structure observed in molecular outflows. 

Unlike other HH jets with a \textit{Spitzer}-identified candidate driving source near the jet axis, [Fe~{\sc ii}] emission from HH~900 does not connect the jet to either of the two YSOs near the globule. 
The jet axis defined by [Fe~{\sc ii}] emission clearly bisects the globule, and dust lanes near the globule surface reveal the impact of the jet breaking out of the globule. 
Together, this argues for the existence of a third protostar that remains deeply embedded in the globule and is obscured by high column density, perhaps by an edge-on circumstellar disc. 
The invisibility of this protostar may be interpreted as evidence that the source is extremely young, that the globule has been compressed to high densities by radiation from nearby massive stars, or both.

Our data allow us to reject the hypothesis that the YSO (PCYC~838) along the western limb of broader bipolar outflow drives a separate microjet. 
The [Fe~{\sc ii}] jet structure is symmetric and center on the globule (not the star) and there is no evidence for multiple velocity components in either H$\alpha$ or [Fe~{\sc ii}] spectra.
Proper motions measured in H$\alpha$ images taken with \textit{HST} $\sim 9$ yr apart are consistent with zero velocity, calling into question the nature of this unusual feature. 
Low velocities rule out a dynamical origin. 
Nevertheless, the small projected separation between HH~900 and this YSO suggests that the YSO may illuminate the nearby bipolar outflow, creating an H$\alpha$-bright filament. 
However, the simplest interpretation is that this YSO is simply a chance alignment with the HH~900 outflow.

The bright, collimated [Fe~{\sc ii}] jet, together with wider-angle H$\alpha$ and H$_2$ components point to a powerful jet breaking out of a small globule that is sufficiently massive to have survived the harsh radiative environment of the Carina nebula. 
Different morphologies and velocity structures from the optical and IR emission lines in HH~900 reveal the protostellar jet itself and offer a rare glimpse of the material it entrains before it is destroyed in the H~{\sc ii} region. 
This supports the interpretation that the irradiated outflows in the Carina reflect the same underlying phenomena driving molecular outflows in embedded regions, but that they are laid bare and illuminated by the harsh radiative environment.


\section*{Acknowledgments}
We would like to thank Rob Simcoe for his assistance with reduction of the FIRE data and Jayadev Rajagopal for assistance with the Gemini Observations.  
MR would like to thank Kaitlin Kratter for helpful discussions. 
Support for this work was provided by NASA through grants AR-12155, GO-13390, and GO-13391 from the Space Telescope Science Institute. 
This work is based on observations made with the NASA/ESA Hubble Space Telescope, obtained from the Data Archive at the Space Telescope Science Institute, which is operated by the Association of Universities for Research in Astronomy, Inc., under NASA contract NAS 5-26555. These \textit{HST} observations are associated with programs GO~10241, 10475, 13390, and 13391. 
Gemini observations are from the GS-2013A-Q-12 science program.



\label{lastpage}

\end{document}